\documentclass[12pt]{article}
\usepackage[dvips]{graphics,color}
\usepackage{latexsym,amssymb}
\usepackage{amsmath,amsbsy}
\usepackage[all]{xy}
\usepackage{amsopn,amstext}
\usepackage{cite}
\usepackage{amssymb}
\usepackage{rotating}
\usepackage{amsmath}
\usepackage[all]{xy}

\allowdisplaybreaks[4]

\begin{document}

%************************** Text Begins here ******************************

%  Greek letters

\def\a{\alpha}
\def\b{\beta}
\def\d{\delta}
\def\e{\epsilon}
\def\g{\gamma}
\def\h{\mathfrak{h}}
\def\k{\kappa}
\def\l{\lambda}
\def\o{\omega}
\def\p{\wp}
\def\r{\rho}
\def\t{\tau}
\def\s{\sigma}
\def\z{\zeta}
\def\x{\xi}
\def\V={{{\bf\rm{V}}}}
 \def\A{{\cal{A}}}
 \def\B{{\cal{B}}}
 \def\C{{\cal{C}}}
 \def\D{{\cal{D}}}
\def\G{\Gamma}
\def\K{{\cal{K}}}
\def\O{\Omega}
\def\R{\bar{R}}
\def\T{{\cal{T}}}
\def\L{\Lambda}
\def\f{E_{\tau,\eta}(sl_2)}
\def\E{E_{\tau,\eta}(sl_n)}
\def\Zb{\mathbb{Z}}
\def\Cb{\mathbb{C}}

\def\R{\overline{R}}
% Shorthands for \begin{equation} and the like

\def\beq{\begin{equation}}
\def\eeq{\end{equation}}
\def\bea{\begin{eqnarray}}
\def\eea{\end{eqnarray}}
\def\ba{\begin{array}}
\def\ea{\end{array}}
\def\no{\nonumber}
\def\le{\langle}
\def\re{\rangle}
\def\lt{\left}
\def\rt{\right}

\newtheorem{Theorem}{Theorem}
\newtheorem{Definition}{Definition}
\newtheorem{Proposition}{Proposition}
\newtheorem{Lemma}{Lemma}
\newtheorem{Corollary}{Corollary}
\newcommand{\proof}[1]{{\bf Proof. }
        #1\begin{flushright}$\Box$\end{flushright}}

\baselineskip=20pt

%%%%%%%%%%%%%%%%%%%%%%%%%%%%%%%%%%%%%%%%%%%%%%%%%%%%%%%%%%%%
%                                                          %
%  Title page                                              %
%                                                          %
%%%%%%%%%%%%%%%%%%%%%%%%%%%%%%%%%%%%%%%%%%%%%%%%%%%%%%%%%%%%
\newfont{\elevenmib}{cmmib10 scaled\magstep1}
\newcommand{\preprint}{
   \begin{flushleft}
     %\elevenmib Yukawa\, Institute\, Kyoto\\
   \end{flushleft}\vspace{-1.3cm}
   \begin{flushright}\normalsize
  % \sf  YITP-03-53\\
   %  {\tt hep-th/yymmnnn} \\ November 2005
   \end{flushright}}
\newcommand{\Title}[1]{{\baselineskip=26pt
   \begin{center} \Large \bf #1 \\ \ \\ \end{center}}}

\newcommand{\Author}{\begin{center}
   \large \bf
Guang-Liang Li${}^{a, b}$,  Panpan
Xue${}^{a}$, Pei Sun${}^{c,d}$, Hulin Yang${}^{a}$, Xiaotian Xu${}^{b,c,d}\footnote{Corresponding author:
xtxu@nwu.edu.cn}$, Junpeng Cao${}^{b, e, f, g}$, Tao Yang${}^{b,c,d}$ and Wen-Li
Yang${}^{b,c,d,h}$
 \end{center}}

\newcommand{\Address}{\begin{center}

     ${}^a$ School of  Physics, Xi'an Jiaotong University, Xian 710049, China\\
    ${}^b$ Peng Huanwu Center for Fundamental Theory,
Xian 710127, China\\
${}^c$ Institute of Modern Physics, Northwest University,
     Xian 710127, China\\
     ${}^d$ Shaanxi Key Laboratory for Theoretical Physics Frontiers,  Xian 710127, China\\
     ${}^e$ Beijing National Laboratory for Condensed Matter
           Physics, Institute of Physics, Chinese Academy of Sciences, Beijing
           100190, China\\
     ${}^f$ School of Physical Sciences, University of Chinese Academy of
Sciences, Beijing, China\\
     ${}^g$ Songshan Lake Materials Laboratory, Dongguan, Guangdong 523808, China \\
          ${}^h$ School of Physics, Northwest University,  Xian 710127, China
\end{center}}

\newcommand{\Accepted}[1]{\begin{center}
   {\large \sf #1}\\ \vspace{1mm}{\small \sf Accepted for Publication}
   \end{center}}

\preprint
\thispagestyle{empty}
\bigskip\bigskip\bigskip

\Title{Exact solutions of the $C_n$ quantum spin chain} \Author

\Address
\vspace{1cm}

\begin{abstract}
\bigskip
We study the exact solutions of quantum integrable model associated with the $C_n$ Lie algebra, with either a periodic or
an open one with off-diagonal boundary reflections, by generalizing the nested off-diagonal Bethe ansatz method.
Taking the $C_3$ as an example we demonstrate how the generalized method works.
We give the fusion structures of the model and
provide a way to close fusion processes. Based on the resulted operator product identities among fused transfer matrices
and some necessary additional constraints such as
asymptotic behaviors and relations at some special points,
we obtain the eigenvalues of transfer matrices
and parameterize them as homogeneous $T-Q$ relations in the periodic case or inhomogeneous ones in
the open case.
We also give the exact solutions of the $C_n$ model with an off-diagonal open boundary condition.
The method and results in this paper can be generalized to other high rank integrable models associated with other Lie algebras.

\vspace{1truecm} \noindent {\it PACS:} 75.10.Pq, 02.30.Ik, 71.10.Pm

\end{abstract}
\newpage
%%%%%%%%%%%%%%%%%%%%%%%%%%%%%%%%%%%%%%%%%%%%%%%%%%%%%%%%%%%%%%%
%                                                             %
%  1. Introduction                                            %
%                                                             %
%%%%%%%%%%%%%%%%%%%%%%%%%%%%%%%%%%%%%%%%%%%%%%%%%%%%%%%%%%%%%%%

\section{Introduction}

Quantum integrable models have many applications in the
fields of quantum field theory, condensed matter physics, string theory and
mathematical physics. The algebraic/coordinate Bethe ansatz and $T-Q$ relations are the very powerful methods to obtain exact solutions of
integrable models with periodic or diagonal open boundary conditions \cite{1,2,Tak79,Alc87,Skl88}.
Focusing on the boundary integrable models, it is well-known that some reflection matrices including the off-diagonal elements also satisfy the reflection equations, which implies
that the systems are still integrable even with off-diagonal boundary reflections. However, due to the existence of off-diagonal elements,
it is quite difficult to calculated the exact solutions of this kind of systems because that the reflection matrices at two boundaries cannot be diagonalized simultaneously.
We also note that the models with off-diagonal boundary reflections are very important and have many applications in many issues such as the open AdS/CFT theory, edge states and topological physics.
Therefore, many interesting methods such as the q-Qnsager algebra method \cite{Bas13,Bas14}, the separation of variables \cite{Fra08, Fra11, Nic13},
the modified algebraic Bethe ansatz \cite{Bel13, Bel15, Pim15, Ava15} and the off-diagonal Bethe ansatz (ODBA) \cite{cao13, wang15} are proposed to study this kind of systems.

The ODBA is an universal method to solve the models with generic integrable boundary conditions. With the help of the proposed inhomogeneous $T-Q$ relations, exact solutions of some typical models with off-diagonal boundary reflections are obtained \cite{wang15}.
Furthermore, in order to solve the models with high ranks \cite{Perk81,NYRes,NYReshetikhin2,Bn,s3,s2,s1}, the nested ODBA has been proposed and
the exact solutions of models
associated with $A_n$ \cite{Cao14, Cao15}, $A_2^{(2)}$ \cite{Hao14}, $B_2$ \cite{Li_119}, $C_2$ \cite{Li_219} and $D_3$ \cite{Li_119} Lie algebras were obtained.
One important property of high rank integrable models is that
the eigenvalue of transfer matrix is a polynomial where the degree is higher, thus we need more functional relations to determine it completely.
Meanwhile, due to the different algebraic structures, the closing conditions of these functional relations are quite different.

In this paper, we study the functional relations of the integrable $C_n$ vertex model by using the fusion technique \cite{Kul81, Kul82,Kul86,Kar79,Kir86,Kir87,Mez92,Zho96} and the nested ODBA \cite{wang15}. Firstly taking the $C_3$ model as an example,
we systemically analyze the fusion behaviors and obtain recursive fusion relations among the fused transfer matrices.
The fusion relations with periodic boundary conditions are different from those with open boundaries.
We provide a way to close these recursive fusion relations.
Based on them and asymptotic behaviors as well as values at certain points, we obtain
the eigenvalues of transfer matrices and parameterize them as the homogeneous or inhomogeneous $T-Q$ relations.
The associated Bethe ansatz equations are also given. Then we generalize these results to the $C_n$ model with off-diagonal open boundary condition.
We expect that the method and results provided in this work can be applied to other
high rank integrable models associated with other Lie algebras.

The plan of the paper is as follows. In section 2, we study the model with periodic boundary condition. The fusion structures of integrable $C_3$ vertex model is shown in detailed.
The closed recursive fusion relations among fused transfer matrices are given. By constructing the $T-Q$ relations, we obtain the eigenvalues and associated Bethe ansatz equations of
the system.
In section 3, we diagonalize the model with off-diagonal boundary reflections. The reflection matrices with off-diagonal elements and corresponding fusion behavior are introduced.
Based on the closed operators product identities, we obtain the eigenvalues
of transfer matrices and expressed them as the inhomogeneous $T-Q$ relations.
These results are also generalized to the $C_n$ model, which are listed in section 4. The summary of main results and some concluding remarks are presented in section 5.

\section{$C_3$ model with periodic boundary condition}
\setcounter{equation}{0}

\subsection{Integrability}

Through this paper, we adopt following standard
notations. Let ${ V}$ denote a $6$-dimensional linear space with
orthogonal bases $\{|i\rangle|i=1,\cdots,6\}$. For any matrix $A\in {\rm End}({ V})$, $A_j$ is an
embedding operator in the tensor space ${ V}\otimes {
V}\otimes\cdots$, which acts as $A$ on the $j$-th space and as an
identity on the other factor spaces. For a matrix $R\in {\rm
End}({ V}\otimes { V})$, $R_{ij}$ is an embedding operator defined in the same tensor space, which acts as an identity on the
factor spaces except for the $i$-th and $j$-th ones.

The quantum integrable system associated with $C_3$ Lie algebra is described by a
$36\times 36$ $R$-matrix $R_{12}(u)$ with the elements \cite{Bn}
 \begin{eqnarray}
R_{12}(u)^{ij}_{kl} = u(u+4)\delta_{ik}\delta_{jl}+
(u+4)\delta_{il}\delta_{jk}-u\xi_i\xi_k
\delta_{j\bar{i}}\delta_{k\bar{l}},
 \label{rm}
 \end{eqnarray}
where $u$ is the spectral parameter, $i+\bar{i}=7$, $\xi_i=1$ if $i\in[1,3]$ while $\xi_i=-1$ if
$i\in[4,6]$. For the simplicity, we introduce following notations
\begin{eqnarray}
&&a(u)=R(u)^{ii}_{ii}=(1+u)(u+4),\quad b(u)=R(u)^{ij}_{ij}=u(u+4),\ \ (i\ne j, \bar{j}),\nonumber\\
&&c(u)=R(u)^{i\bar{i}}_{\bar{i}i}=2u+4,\quad
d(u)=\xi_i\xi_jR(u)^{i\bar{i}}_{j\bar{j}}=-u,\ \ (i\ne j, \bar{j}),\nonumber\\
&&e(u)=R(u)^{i\bar{i}}_{i\bar{i}}=u(u+3),\quad
g(u)=R(u)^{ij}_{j{i}}=u+4,\ \ (i\ne j, \bar{j}).
 \label{2}
\end{eqnarray}
The $R$-matrix (\ref{rm}) has following properties
\begin{eqnarray}
{\rm regularity}&:&R_{12}(0)=\rho_v(0)^{\frac{1}{2}}{\cal P}_{12},\nonumber\\
{\rm unitarity}&:&R_{12}(u)R_{21}(-u)=\rho_v(u),\nonumber\\
{\rm crossing-unitarity}&:&R_{12}(u)^{t_1}R_{21}(-u-8)^{t_1}=\tilde{\rho}_v(u)=\rho_v(u+4),
\end{eqnarray}
where $\rho_v(u)=a(u)a(-u)$, ${\cal P}_{12}$ is the permutation
operator with the matrix elements $[{\cal
P}_{12}]^{ij}_{kl}=\delta_{il}\delta_{jk}$, $t_i$ denotes the
transposition in the $i$-th space, and $R _{21}={\cal P}_{12}R
_{12}{\cal P}_{12}$. The $R$-matrix (\ref{rm}) satisfies the Yang-Baxter equation
\begin{eqnarray}
R_{12}(u-v)R_{13}(u)R_{23}(v)=R_{23}(v)R_{13}(u)R_{12}(u-v). \label{YBE}
\end{eqnarray}

The monodromy matrix of the system is constructed by the $R$-matrix (\ref{rm}) as
\bea
T_0(u)=R_{01}(u-\theta_1)R_{02}(u-\theta_2)\cdots R_{0N}(u-\theta_N), \label{Mon-1}
\eea
where the subscript $0$ means the auxiliary space,
the other tensor space  $V^{\otimes N}$ is the physical or quantum space, $N$ is the number of sites and $\{\theta_j|j=1,\cdots,N\}$ are the inhomogeneous parameters.
The monodromy matrix satisfies the Yang-Baxter relation
\bea
 R_{12}(u-v) T_1(u) T_2(v) = T_2(v) T_1(u) R_{12}(u-v).\label{ybta2o}
\eea
Taking the partial trace of monodromy matrix in the
auxiliary space, we arrive at the transfer matrix of the system with periodic boundary condition
\bea t^{(p)}(u)=tr_0 T_0(u). \label{1117-1}\eea From the
Yang-Baxter relation (\ref{ybta2o}), one can prove that the transfer matrices
with different spectral parameters commutate with each other, i.e.,
$[t^{(p)}(u), t^{(p)}(v)]=0$. Therefore, $t^{(p)}(u)$ serves as
the generating function of all the conserved quantities of the
system. The model Hamiltonian with $C_3$-invariant is given by
\begin{eqnarray}
H_p= \frac{\partial \ln t^{(p)}(u)}{\partial
u}|_{u=0,\{\theta_j\}=0}. \label{asdasd}
\end{eqnarray}

\subsection{Fusion}

One wonderful property of $R$-matrix is that the $R$-matrix may degenerate into the projection
operators at some special points, which makes it possible for us to do the fusion.
Focus on the $C_3$ model, the elements of $R$-matrix (\ref{rm}) are the polynomials of $u$ with degree two. Thus there are two degenerate points. One is $u=-4$. At which we have
\bea
R_{12}(-4)=  P^{(1) }_{12}\times S^\prime_{12}.\label{Int-R1}
\eea
Here $P^{ (1) }_{12}$ is a one-dimensional projection operator
with the form
\bea
P^{(1)}_{12}=|\psi_0\rangle \langle \psi_0|,\label{1-project}
\eea
where
$|\psi_0\rangle=\frac{1}{\sqrt{6}}(|16\rangle+|25\rangle+|34\rangle-|43\rangle-|52\rangle-|61\rangle)$
is a one-dimensional vector in the product space ${V_1}\otimes {V_2}$ and $S^\prime_{12}$ is a
constant matrix (we omit its expression because we do not need
it). Obviously, $P^{(1) }_{21}=P^{(1) }_{12}$.
From the Yang-Baxter equation (\ref{YBE}), the one-dimensional fusion associated with projector (\ref{1-project}) leads to
\bea &&P^{(1) }_{21}R_{13}(u)R_{23}(u-4)P^{(1)}_{21}=a(u)e(u-4)P^{(1) }_{21}\times {\rm
id}.\label{Quantum-det} \eea
We see that the result is also a one-dimensional vector.

The other degenerate point of $R$-matrix (\ref{rm}) is $u=-1$. At which we have
\bea
R_{12}(-1)=  P_{12}^{(14)}\times
S_{12}.\label{Int-R2}
\eea
Here $S_{12}$ is a constant matrix and $P^{(14)}_{12}$ is a
14-dimensional projection operator with the form of
\bea
P^{(14) }_{12}=\sum_{i=1}^{14} |{\psi}^{(14)}_i\rangle \langle {\psi}^{(14)}_i|, \quad P^{(14) }_{21}=P^{(14) }_{12},
\label{2-project}\eea where the corresponding
vectors are
\bea
&&|{\psi}^{(14)}_1\rangle=\frac{1}{\sqrt{2}}(|12\rangle-|21\rangle),\quad |{\psi}^{(14)}_2\rangle=\frac{1}{\sqrt{2}}(|13\rangle-|31\rangle),\quad |{\psi}^{(14)}_3\rangle=\frac{1}{\sqrt{2}}(|14\rangle-|41\rangle),\nonumber\\
&&|{\psi}^{(14)}_4\rangle=\frac{1}{\sqrt{2}}(|15\rangle-|51\rangle),\quad |{\psi}^{(14)}_5\rangle=\frac{1}{{2}}(|16\rangle-|61\rangle+|43\rangle-|34\rangle),\nonumber\\
&&|{\psi}^{(14)}_6\rangle=\frac{1}{\sqrt{2}}(|23\rangle-|32\rangle),\quad |{\psi}^{(14)}_7\rangle=\frac{1}{\sqrt{2}}(|24\rangle-|42\rangle),\nonumber\\
&&|{\psi}^{(14)}_8\rangle=\frac{1}{\sqrt{12}}(-|16\rangle+|61\rangle+|43\rangle-|34\rangle+2|25\rangle-2|52\rangle),\nonumber\\
&&|{\psi}^{(14)}_9\rangle=\frac{1}{\sqrt{2}}(|26\rangle-|62\rangle),\quad |{\psi}^{(14)}_{10}\rangle=\frac{1}{\sqrt{2}}(|35\rangle-|53\rangle),\quad |{\psi}^{(14)}_{11}\rangle=\frac{1}{\sqrt{2}}(|36\rangle-|63\rangle),\nonumber\\
&&|{\psi}^{(14)}_{12}\rangle=\frac{1}{\sqrt{2}}(|45\rangle-|54\rangle),\quad |{\psi}^{(14)}_{13}\rangle=\frac{1}{\sqrt{2}}(|46\rangle-|64\rangle),\quad |{\psi}^{(14)}_{14}\rangle=\frac{1}{\sqrt{2}}(|56\rangle-|65\rangle).
\no\eea
From the 14-dimensional fusion associated with the projector
(\ref{2-project}), we obtain a new fused $R$-matrix
\bea
R_{\langle 12\rangle 3}(u)
=\tilde{\rho}_0^{-1}(u+\frac{1}{2})P^{(14)}_{21}R_{13}(u+\frac{1}{2})
R_{23}(u-\frac{1}{2})P^{(14)}_{21}\equiv  R_{\bar{1} 3}(u),\label{Fused-R-1} \eea
where $\tilde{\rho}_0(u)=(u-1)(u+4)$.
We note that the dimension of fused space $V_{\langle 12\rangle}=V_{\bar 1}$ is 14. The fused $R$-matrix (\ref{Fused-R-1}) has the properties
\begin{eqnarray}
&&R_{\bar{1}2}(u)R_{2\bar{1}}(-u)={\rho}_{\bar{v}}(u)\times{\rm id}, \no  \\
&&R_{\bar{1}2}(u)^{t_{\bar{1}}}R_{2\bar{1}}(-u-8)^{t_{\bar{1}}}=\tilde{\rho}_{\bar{v}}(u)\times{\rm id}, \no \\
&&R_{\bar{1}2}(u-v)R_{\bar{1}3}(u)R_{23}(v)=R_{23}(v)R_{\bar{1}3}(u)R_{\bar{1}2}(u-v),\label{lic-13}
\end{eqnarray}
where ${\rho}_{\bar{v}}(u)=(u+\frac{7}{2})(u-\frac{7}{2})(u+\frac{3}{2})(u-\frac{3}{2})$ and $\tilde{\rho}_{\bar{v}}(u)=(u+\frac{1}{2})(u+\frac{5}{2})(u+\frac{11}{2})(u+\frac{15}{2})$.

The elements of fused $R$-matrix (\ref{Fused-R-1}) are the polynomials of $u$ with degree two. Thus there are two degenerate points. One is
$u=- 7/2$, at which the fused $R$-matrix
$R_{\bar{1}2}(u)$ degenerates into a 6-dimensional projector
\bea R_{\bar{1}2}(-\frac{7}{{2}})= P^{(6) }_{\bar{1}2}\times
S_{\bar{1}2},\label{pbv} \eea where $S_{\bar{1}2}$ is not relevant here and we do not
present its expression for simplicity, $P^{(6)
}_{\bar{1}2}$ is a $6$-dimensional projector
\bea
&& P^{(6) }_{\bar{1}2}=\sum_{i=1}^{6} |{\psi}^{(6)}_i\rangle \langle {\psi}^{(6)}_i|, \label{6bar-project}
\eea
and the corresponding bases are
\bea
&&|{\psi}^{(6)}_1\rangle=\sqrt{\frac{3}{14}}(-|\bar 15\rangle-|\bar 24\rangle+|\bar 33\rangle+|\bar 42\rangle+\sqrt{\frac{1}{2}}|\bar 51\rangle-\sqrt{\frac{1}{6}}|\bar 81\rangle),\nonumber\\
&&|{\psi}^{(6)}_2\rangle=\sqrt{\frac{3}{14}}(|\bar 16\rangle-|\bar 64\rangle+|\bar 73\rangle+|\bar 91\rangle+\sqrt{\frac{2}{3}}|\bar 82\rangle),\nonumber\\
&&|{\psi}^{(6)}_3\rangle=\sqrt{\frac{3}{14}}(|\bar 26\rangle+|\bar 65\rangle+|\bar {10}, 2\rangle+|\bar {11}, 1\rangle-\sqrt{\frac{1}{2}}|\bar 53\rangle-\sqrt{\frac{1}{6}}|\bar 83\rangle),\nonumber\\
&&|{\psi}^{(6)}_4\rangle=\sqrt{\frac{3}{14}}(|\bar 36\rangle+|\bar 75\rangle+|\bar {12}, 2\rangle+|\bar {13}, 1\rangle-\sqrt{\frac{1}{2}}|\bar 54\rangle-\sqrt{\frac{1}{6}}|\bar 84\rangle),\nonumber\\
&&|{\psi}^{(6)}_5\rangle=\sqrt{\frac{3}{14}}(|\bar 46\rangle+|\bar {10}, 4\rangle-|\bar {12},3\rangle+|\bar {14}, 1\rangle+\sqrt{\frac{2}{3}}|\bar 85\rangle),\nonumber\\
&&|{\psi}^{(6)}_6\rangle=\sqrt{\frac{3}{14}}(|\bar 95\rangle+|\bar {11}, 4\rangle-|\bar {13},3\rangle-|\bar {14},
2\rangle+\sqrt{\frac{1}{2}}|\bar 56\rangle-\sqrt{\frac{1}{6}}|\bar 86\rangle).\no \eea
The projector $P^{(6) }_{2\bar{1}}$ can be obtained from $P^{(6) }_{\bar{1}2}$ by exchanging the bases of $V_{\bar 1}$ and $V_2$.
The projector (\ref{6bar-project}) shows that we can fuse the spaces $V_{\bar 1}$ and $V_2$, and the result is that we obtain
a new fused
$R$-matrix,
\bea R_{\langle\bar{1}2\rangle3}(u)=\tilde{\rho}_0^{-1}(u+3) P^{(6) }_{\bar{1}2}R_{23}(u+3)R_{\bar{1}3}(u-\frac{1}{2})P^{(6)}_{\bar{1}2}. \label{6r} \eea
We note the dimension of the fused space $V_{\langle\bar{1}2\rangle}$ is 6. Thus fused $R$ matrix (\ref{6r}) is a $36\times 36$ one.
Taking the correspondence \bea
|\psi^{(6)}_i\rangle\longrightarrow |i\rangle,\quad
i=1,\cdots,6,\label{Identification-1} \eea
we find that the fused $R$-matrix (\ref{6r}) is the same as the original one (\ref{rm}), i.e.,
\bea R_{\langle\bar{1}2\rangle3}(u)= R_{13}(u).\label{R5-5} \eea

We remark that from the way of above fusion, the auxiliary space cannot be enlarged anymore. However, both the
orders of elements of $R$-matrix (\ref{rm}) and that of the fused one (\ref{Fused-R-1}) are two. Therefore, the above fusion processes indeed are not closed and we should go further.

In order to obtain the closed fusion relations among fused $R$-matrices,
we have to consider the degenerations of fused $R$-matrix (\ref{Fused-R-1}) at the other degenerate point, $u=-3/2$. At which, the
fused $R$-matrix (\ref{Fused-R-1}) has a 14-dimensional projected subspace, which can be seen from the identity
\bea R_{\bar{1}2}(-\frac{3}{{2}})= P^{(14) }_{\bar{1}2}\times
S_{\bar{1}2}^\prime,\no \eea where $S_{\bar{1}2}^\prime$ is an irrelevant constant matrix,
$P^{(14) }_{\bar{1}2}$ is the $14$-dimensional projector \bea &&
P^{(14) }_{\bar{1}2}=\sum_{i=1}^{14}
|{\bar{\psi}}^{(14)}_i\rangle \langle{\bar{\psi}}^{(14)}_i|,\label{new-14-pro} \eea
and the corresponding bases are
\bea
&&|{\bar{\psi}}^{(14)}_1\rangle=\frac{1}{\sqrt{3}}(|\bar 13\rangle-|\bar 22\rangle+|\bar 61\rangle),
\quad |{\bar{\psi}}^{(14)}_2\rangle=\frac{1}{\sqrt{3}}(|\bar 14\rangle-|\bar 32\rangle+|\bar 71\rangle),\nonumber\\
&&|{\bar{\psi}}^{(14)}_3\rangle=\frac{1}{\sqrt{10}}(\sqrt{2}|\bar 24\rangle
-\sqrt{2}|\bar 33\rangle+\sqrt{2}|\bar 42\rangle-|\bar 51\rangle-\sqrt{3}|\bar 81\rangle),\nonumber\\
&&|{\bar{\psi}}^{(14)}_4\rangle=\frac{1}{\sqrt{2}}(|\bar 43\rangle-|\bar {10},1\rangle),\quad |{\bar{\psi}}^{(14)}_5\rangle=\frac{1}{\sqrt{2}}(|\bar 44\rangle-|\bar {12},1\rangle),\nonumber\\
&&|{\bar{\psi}}^{(14)}_6\rangle=\frac{1}{\sqrt{5}}(\sqrt{2}|\bar 52\rangle-|\bar 64\rangle+|\bar 73\rangle-|\bar 91\rangle),\nonumber\\
&&|{\bar{\psi}}^{(14)}_7\rangle=\frac{1}{\sqrt{8}}(|\bar 53\rangle-\sqrt{3}|\bar 83\rangle+\sqrt{2}|\bar {10},2\rangle-\sqrt{2}|\bar {11},1\rangle),\nonumber\\
&&|{\bar{\psi}}^{(14)}_8\rangle=\frac{1}{\sqrt{8}}(|\bar 54\rangle-\sqrt{3}|\bar 84\rangle+\sqrt{2}|\bar {12},2\rangle-\sqrt{2}|\bar {13},1\rangle),\nonumber\\
&&|{\bar{\psi}}^{(14)}_9\rangle=\frac{1}{\sqrt{2}}(|\bar 93\rangle-|\bar {11},2\rangle),\quad |{\bar{\psi}}^{(14)}_{10}\rangle=\frac{1}{\sqrt{2}}(|\bar 94\rangle-|\bar {13},2\rangle),\nonumber\\
&&|{\bar{\psi}}^{(14)}_{11}\rangle=\frac{1}{\sqrt{3}}(|\bar {10},4\rangle-|\bar {12},3\rangle-|\bar {14},1\rangle),\quad |{\bar{\psi}}^{(14)}_{12}\rangle=\frac{1}{\sqrt{3}}(|\bar {11},4\rangle-|\bar {13},3\rangle-|\bar {14},2\rangle),\nonumber\\
&&|{\bar{\psi}}^{(14)}_{13}\rangle=|\bar {14},3\rangle,\quad |{\bar{\psi}}^{(14)}_{14}\rangle=|\bar {14},4\rangle. \no \eea
It is obvious that the projector $P^{(14) }_{2\bar{1}}$ can be obtained from $P^{(14) }_{\bar{1}2}$ by exchanging the bases of $V_{\bar 1}$ and $V_2$.
The projector (\ref{new-14-pro}) is survived in the tensor space $V_1\otimes V_2\otimes V_3$. By carefully analyzing the fusion structure,
We find that the 14-dimensional projected space defined by (\ref{new-14-pro}) can also be obtained from the product of three $R$-matrices (\ref{rm}) at certain points with the following way
\bea
R_{12}(-1)R_{13}(-2)R_{23}(-1)= P_{123}^{(14)}\times
S_{123},\label{Int-R3}
\eea
where $S_{123}$ is constant matrix, $P_{123}^{(14)}$ is a 14-dimensional projector defined in the spaces $V_1\otimes V_2\otimes V_3$
\bea
P^{(14) }_{123}=\sum_{i=1}^{14}
|{\phi}^{(14)}_i\rangle \langle {\phi}^{(14)}_i|,\quad P^{(14) }_{321}=P^{(14) }_{123}\label{2-14-project}
\eea
and the corresponding bases are
\bea
&&|{\phi}^{(14)}_1\rangle=\frac{1}{\sqrt{6}}(|123\rangle-|132\rangle-|213\rangle+|231\rangle+|312\rangle-|321\rangle),\nonumber\\
&&|{\phi}^{(14)}_2\rangle=\frac{1}{\sqrt{6}}(|124\rangle-|142\rangle-|214\rangle+|241\rangle+|412\rangle-|421\rangle),\nonumber\\
&&|{\phi}^{(14)}_3\rangle=\frac{1}{\sqrt{12}}(|125\rangle-|152\rangle-|215\rangle+|251\rangle+|512\rangle-|521\rangle\no\\
&&\hspace{25mm}-|134\rangle+|143\rangle+|314\rangle-|341\rangle-|413\rangle+|431\rangle),\nonumber\\
&&|{\phi}^{(14)}_4\rangle=\frac{1}{\sqrt{12}}(|126\rangle-|162\rangle-|216\rangle+|261\rangle+|612\rangle-|621\rangle\no\\
&&\hspace{25mm}+|234\rangle-|243\rangle-|324\rangle+|342\rangle+|423\rangle-|432\rangle),\nonumber\\
&&|{\phi}^{(14)}_5\rangle=\frac{1}{\sqrt{6}}(|135\rangle-|153\rangle-|315\rangle+|351\rangle+|513\rangle-|531\rangle),\nonumber\\
&&|{\phi}^{(14)}_6\rangle=\frac{1}{\sqrt{12}}(|136\rangle-|163\rangle-|316\rangle+|361\rangle+|613\rangle-|631\rangle\no\\
&&\hspace{25mm}-|235\rangle+|253\rangle+|325\rangle-|352\rangle-|523\rangle+|532\rangle),\nonumber\\
&&|{\phi}^{(14)}_7\rangle=\frac{1}{\sqrt{6}}(|145\rangle-|154\rangle-|415\rangle+|451\rangle+|514\rangle-|541\rangle),\nonumber\\
&&|{\phi}^{(14)}_8\rangle=\frac{1}{\sqrt{12}}(|146\rangle-|164\rangle-|416\rangle+|461\rangle+|614\rangle-|641\rangle\no\\
&&\hspace{25mm}-|245\rangle+|254\rangle+|425\rangle-|452\rangle-|524\rangle+|542\rangle),\nonumber\\
&&|{\phi}^{(14)}_9\rangle=\frac{1}{\sqrt{12}}(|156\rangle-|165\rangle-|516\rangle+|561\rangle+|615\rangle-|651\rangle\no\\
&&\hspace{25mm}+|345\rangle-|354\rangle-|435\rangle+|453\rangle+|534\rangle-|543\rangle),\nonumber\\
&&|{\phi}^{(14)}_{10}\rangle=\frac{1}{\sqrt{6}}(|236\rangle-|263\rangle-|326\rangle+|362\rangle+|623\rangle-|632\rangle),\nonumber\\
&&|{\phi}^{(14)}_{11}\rangle=\frac{1}{\sqrt{6}}(|246\rangle-|264\rangle-|426\rangle+|462\rangle+|624\rangle-|642\rangle),\nonumber\\
&&|{\phi}^{(14)}_{12}\rangle=\frac{1}{\sqrt{12}}(|256\rangle-|265\rangle-|526\rangle+|562\rangle+|625\rangle-|652\rangle\no\\
&&\hspace{25mm}-|346\rangle+|364\rangle+|436\rangle-|463\rangle-|634\rangle+|643\rangle),\nonumber\\
&&|{\phi}^{(14)}_{13}\rangle=\frac{1}{\sqrt{6}}(|356\rangle-|365\rangle-|536\rangle+|563\rangle+|635\rangle-|653\rangle),\no\\
&&|{\phi}^{(14)}_{14}\rangle=\frac{1}{\sqrt{6}}(|456\rangle-|465\rangle-|546\rangle+|564\rangle+|645\rangle-|654\rangle).\no
\eea
We note that the projectors (\ref{new-14-pro}) and (\ref{2-14-project}) give the same subspace, and the only difference is the bases. By taking the
suitable gauge transformation, we can map the projector (\ref{new-14-pro}) into (\ref{2-14-project}) and vice versa.
Thus the finial results of fusion are equivalent. Here we only give the fusion results with the projector (\ref{2-14-project}).

Taking the fusion with projector (\ref{2-14-project}), we construct another fused $R$-matrix
\bea
&&R_{\langle 123\rangle 4}(u) =[\tilde{\rho}_0(u+1)\tilde{\rho}_0(u)(u+2)]^{-1}
P_{321}^{(14)}R_{14}(u+1)R_{24}(u)
R_{34}(u-1)P_{321}^{(14)}\no \\
&&\qquad \qquad \;\;\equiv R_{ \tilde{1} 4}(u).\label{Fused-R-2} \eea We note that
the dimension of the fused space $V_{\langle 123\rangle}=V_{\tilde
1}$ is 14. In the above construction, we have used the relation
(\ref{Int-R3}). The fused $R$-matrix (\ref{Fused-R-2}) has
following properties \bea
&&R_{\tilde{1}2}(u)R_{2\tilde{1}}(-u)={\rho}_{\tilde{v}}(u)\times{\rm id}, \no \\
&&R_{\tilde{1}2}(u)^{t_{\tilde{1}}}R_{2\tilde{1}}(-u-8)^{t_{\tilde{1}}}=\tilde{\rho}_{ \tilde{v}}(u)\times{\rm id}, \no \\
&&R_{\tilde{1}2}(u-v)R_{\tilde{1}3}(u)R_{23}(v)=R_{23}(v)R_{\tilde{1}3}(u)R_{\tilde{1}2}(u-v),\label{lic-23}
\eea
where ${\rho}_{\tilde{v}}(u)=-(u+3)(u-3)$ and $\tilde{\rho}_{\tilde{v}}(u)=-(u+1)(u+7)$.

The elements of fused $R$-matrix (\ref{Fused-R-2}) are the polynomials of $u$ with degree one.
Thus there is only one degenerate point $u=-3$. At which, we have
\bea
R_{\tilde{1}2}(-3)= P^{(14) }_{\tilde{1}2}\times S_{\tilde{1}2}, \label{ptv}
\eea
where $S_{\tilde{1}2}$ is an irrelevant constant matrix omitted here, $P^{(14)}_{\tilde{1}2}$ is a $14$-dimensional
projector
\bea
P^{(14)}_{\tilde{1}2}=\sum_{i=1}^{14}
|{\varphi}^{(14)}_i\rangle \langle {\varphi}^{(14)}_i|,\label{3-14-project} \eea
and the corresponding bases are
\bea
&&|{\varphi}^{(14)}_1\rangle=\sqrt{\frac{1}{6}}(-\sqrt{2}|\tilde 14\rangle+\sqrt{2}|\tilde 23\rangle+|\tilde 32\rangle+|\tilde 41\rangle),\nonumber\\
&&|{\varphi}^{(14)}_2\rangle=\sqrt{\frac{1}{6}}(\sqrt{2}|\tilde 15\rangle+\sqrt{2}|\tilde 52\rangle-|\tilde 33\rangle+|\tilde 61\rangle),\nonumber\\
&&|{\varphi}^{(14)}_3\rangle=\sqrt{\frac{1}{6}}(\sqrt{2}|\tilde 25\rangle+\sqrt{2}|\tilde 72\rangle-|\tilde 34\rangle+|\tilde 81\rangle),\nonumber\\
&&|{\varphi}^{(14)}_4\rangle=\sqrt{\frac{1}{6}}(\sqrt{2}|\tilde 54\rangle-\sqrt{2}|\tilde 73\rangle+|\tilde 35\rangle+|\tilde 91\rangle),\nonumber\\
&&|{\varphi}^{(14)}_5\rangle=\sqrt{\frac{1}{12}}({2}|\tilde 45\rangle-{2}|\tilde 92\rangle-|\tilde 36\rangle+|\tilde 64\rangle-|\tilde 83\rangle+|\tilde {12},1\rangle),\nonumber\\
&&|{\varphi}^{(14)}_6\rangle=\sqrt{\frac{1}{6}}(-\sqrt{2}|\tilde 16\rangle+\sqrt{2}|\tilde {10},1\rangle+|\tilde 43\rangle-|\tilde 62\rangle),\nonumber\\
&&|{\varphi}^{(14)}_7\rangle=\sqrt{\frac{1}{6}}(-\sqrt{2}|\tilde 26\rangle+\sqrt{2}|\tilde {11},1\rangle+|\tilde 44\rangle-|\tilde 82\rangle),\nonumber\\
&&|{\varphi}^{(14)}_8\rangle={\frac{1}{2}}(-|\tilde 36\rangle-|\tilde 64\rangle+|\tilde 83\rangle+|\tilde {12},1\rangle),\nonumber\\
&&|{\varphi}^{(14)}_9\rangle=\sqrt{\frac{1}{6}}(\sqrt{2}|\tilde {10},4\rangle-\sqrt{2}|\tilde {11},3\rangle-|\tilde 46\rangle-|\tilde {12},2\rangle),\nonumber\\
&&|{\varphi}^{(14)}_{10}\rangle=\sqrt{\frac{1}{6}}(-\sqrt{2}|\tilde 56\rangle+\sqrt{2}|\tilde {13},1\rangle+|\tilde 65\rangle-|\tilde 93\rangle),\nonumber\\
&&|{\varphi}^{(14)}_{11}\rangle=\sqrt{\frac{1}{6}}(-\sqrt{2}|\tilde {10},5\rangle-\sqrt{2}|\tilde {13},2\rangle-|\tilde 66\rangle+|\tilde {12},3\rangle),\nonumber\\
&&|{\varphi}^{(14)}_{12}\rangle=\sqrt{\frac{1}{6}}(-\sqrt{2}|\tilde 76\rangle+\sqrt{2}|\tilde {14},1\rangle+|\tilde 85\rangle-|\tilde 94\rangle),\nonumber\\
&&|{\varphi}^{(14)}_{13}\rangle=\sqrt{\frac{1}{6}}(-\sqrt{2}|\tilde {11},5\rangle-\sqrt{2}|\tilde {14},2\rangle-|\tilde 86\rangle+|\tilde {12},4\rangle),\nonumber\\
&&|{\varphi}^{(14)}_{14}\rangle=\sqrt{\frac{1}{6}}(-\sqrt{2}|\tilde {13},4\rangle+\sqrt{2}|\tilde {14},3\rangle-|\tilde 96\rangle-|\tilde {12},5\rangle).\no
\eea
Again, the projector $P^{(14) }_{2\tilde{1}}$ can be obtained from $P^{(14) }_{\tilde{1}2}$ by exchanging the bases of $V_{\tilde 1}$ and $V_2$.

Taking the fusion of $R$-matrix (\ref{Fused-R-2}) in the auxiliary space by using the 14-dimensional projector $P^{(14) }_{\tilde{1}2}$, we obtain a fused $R$-matrix
\bea
R_{\langle \tilde{1}2\rangle 3}(u)=(u+\frac{13}{2})^{-1} P^{(14) }_{\tilde{1}2}R_{23}(u+\frac{5}{2})R_{\tilde{1}3}(u-\frac{1}{2})P^{(14)}_{\tilde{1}2}. \label{Isd-1} \eea
The dimension of fused space $V_{ \langle \tilde{1}2\rangle}$ is 14, which equals to the dimension of fused space $V_{\bar{1}}$.
After taking the correspondence \bea
|\varphi^{(14)}_i\rangle\longrightarrow
|\psi^{(14)}_i\rangle,\quad i=1,\cdots,14,\label{Identification-2}
\eea
we find the fused $R$-matrix (\ref{Isd-1}) is the same as the fused one (\ref{Fused-R-1}), i.e.,
\bea R_{\langle \tilde{1}2\rangle
3}(u) = R_{\bar{1}3}(u).\label{R11-R11} \eea
Eq.(\ref{R11-R11}) gives another intrinsic relation to close the fusion processes.

Taking the fusion of $R$-matrix (\ref{Fused-R-2}) in the quantum space by using the 14-dimensional projector $P^{(14) }_{23}$ given by (\ref{2-project}), we obtain a fused $R$-matrix
\bea
R_{\tilde{1}\langle
23\rangle }(u)= P^{(14)
}_{23}R_{\tilde{1}2}(u+\frac{1}{2})R_{\tilde{1}3}(u-\frac{1}{2})P^{(14)}_{23}\equiv R_{\tilde{1}\bar{2} }(u).\label{Fuse2d-R-2}
\eea
The fused $R$-matrix (\ref{Fuse2d-R-2}) is defined in the tensor space $V_{\tilde 1}\otimes V_{\bar 2}$ and
has following properties
\bea
&&R_{\tilde{1}\bar{2}}(u)\,R_{\bar{2}\tilde{1}}(-u)={\rho}_{\bar{v}\tilde{v}}(u)\times{\rm id}, \no \\
&&R_{\tilde{1}\bar{2}}(u)^{t_{\bar{2}}}R_{\bar{2}\tilde{1}}(-u-8)^{t_{\bar{2}}}
=\tilde{\rho}_{ \bar{v}\tilde{v}}(u)\times{\rm id}, \no \\
&&R_{\bar{1}\tilde{2}}(u-v)R_{\bar{1}3}(u)R_{\tilde{2}3}(v)=R_{\tilde{2}3}(v)R_{\bar{1}3}(u)R_{\bar{1}\tilde{2}}(u-v).\label{Fuswrqwe2d-R-21}
\eea
where ${\rho}_{\bar{v}\tilde{v}}(u)=(u+\frac{5}{2})(u-\frac{5}{2})(u+\frac{7}{2})
(u-\frac{7}{2})$ and $\tilde{\rho}_{ \bar{v}\tilde{v}}(u)=(u+\frac{1}{2})(u+\frac{3}{2})(u+\frac{13}{2})(u+\frac{15}{2})$.

Last, we remark that the following identity holds
\bea
R_{12}(-1)R_{13}(-2)R_{14}(-3)R_{23}(-1)R_{24}(-2)R_{34}(-1)= 0,\label{Int-R4}
\eea
which can be checked by direct calculation. Eq.(\ref{Int-R4}) implies
that we can not obtain more nontrivial fused $R$-matrix
if we take fusion only in the auxiliary spaces.

\subsection{Operator product identities}

Based on the obtained fused $R$-matrices, we define the fused monodromy matrices
\begin{eqnarray}
 &&T_{\bar{0}}(u)=R_{\bar{0}1}(u-\theta_1)R_{\bar{0}2}(u-\theta_2)\cdots R_{\bar{0}N}(u-\theta_N), \no \\
 &&T_{\tilde{0}}(u)=R_{\tilde{0}1}(u-\theta_1)R_{\tilde{0}2}(u-\theta_2)\cdots R_{\tilde{0}N}(u-\theta_N).  \label{T4}
\end{eqnarray}
We note that the quantum spaces of the above monodromy matrices are the same, which is $V^{\otimes N}$, and the corresponding auxiliary spaces
are $V_{\bar{0}}$ and $V_{\tilde{0}}$ with dimension 14.
The fused monodromy matrices (\ref{T4}) satisfy the
Yang-Baxter relations
\bea
&& R_{1\bar 2}(u-v) T_1(u) T_{\bar 2}(v) = T_{\bar 2}(v) T_1(u) R_{1\bar 2}(u-v), \no \\
&& R_{1\tilde 2}(u-v) T_1(u) T_{\tilde 2}(v) = T_{\tilde 2}(v) T_1(u) R_{1\tilde  2}(u-v), \no \\
&& R_{\bar 1\tilde 2}(u-v) T_{\bar 1}(u) T_{\tilde 2}(v) = T_{\tilde 2}(v) T_{\bar 1}(u) R_{\bar 1\tilde  2}(u-v). \label{ybta2o-2}
\eea
where $R_{\bar 1\tilde  2}(u)$ is the fused $R$-matrix defined in the fused space $V_{\bar 1}\otimes V_{\tilde 2}$, which can be determined by
the first equation in (\ref{Fuswrqwe2d-R-21}).
Besides the transfer matrix $t^{(p)}(u)$, let us introduce two fused transfer matrices
\bea
t^{(p)}_2(u)=tr_{\bar{0}} T_{\bar{0}}(u),\quad t^{(p)}_3(u)=tr_{\tilde{0}} T_{\tilde{0}}(u).\label{Fused-transfer-matrix-periodic}
\eea
From above Yang-Baxter relations (\ref{ybta2o}) and (\ref{ybta2o-2}), we can prove these
transfer matrices commutate with each other, namely, \bea
[{t}^{(p)}(u),t^{(p)}_2(u)]=[{t}^{(p)}(u),t^{(p)}_3(u)]=[{t}^{(p)}_2(u),t^{(p)}_3(u)]=0.
\eea
Therefore, they have common eigenstates and can be diagonalized simultaneously.

By using the above fusion relations of $R$-matrices and the definitions (\ref{Mon-1}) and (\ref{T4}),
we obtain the fusion behavior of monodromy matrices
\bea &&P^{(1)}_{21}T_1(u)T_2(u-4)P^{(1)}_{21}=T_1(u)T_2(u-4)P^{(1)}_{21} \no \\
&& \qquad =\prod_{i=1}^N
a(u-\theta_i)e(u-\theta_i-4)\,P^{(1)}_{21}\times {\rm id}, \no \\
&&P_{21}^{(14)}T_1(u)T_2(u-1)P_{21}^{(14)}=T_1(u)T_2(u-1)P_{21}^{(14)}
\no \\
&& \qquad
= T_{\langle 12\rangle }(u)=\prod_{i=1}^N\tilde{\rho}_0(u-\theta_i)\,
{T}_{\bar 1 }(u-\frac{1}{{2}}), \no \\
&&P_{321}^{(14)}T_1(u)T_2(u-1)T_3(u-2)P_{321}^{(14)}=T_1(u)T_2(u-1)T_3(u-2)P_{321}^{(14)}
\no \\
&&\qquad =\prod_{i=1}^N
\tilde{\rho}_0(u-\theta_i)\tilde{\rho}_0(u-\theta_i-1)(u-\theta_i+1)\,{T}_{\tilde
1 }(u-1), \no \\
&&P^{(6) }_{\bar{1}2}{T}_2(u){T}_{\bar{1}}(u-\frac{7}{{2}})P^{(6) }_{\bar{1}2}={T}_2(u){T}_{\bar{1}}(u-\frac{7}{{2}})P^{(6) }_{\bar{1}2}
=\prod_{i=1}^N
\tilde{\rho}_0(u+\theta_i){T}_{1}(u-3), \no \\[4pt]
&&P^{(14) }_{\tilde{1}2}{T}_2(u){T}_{\tilde{1}}(u-3)P^{(14)
}_{\tilde{1}2}={T}_2(u){T}_{\tilde{1}}(u-3)P^{(14)}
=\prod_{i=1}^N (u+\theta_i+4){T}_{\bar 1}(u-\frac{5}{{2}}).\label{fu11tt-7}
\eea
Here the subscripts $1$ and $2$ mean the original 6-dimensional auxiliary spaces $V_1$ and $V_2$,
the $\bar 1$ means the 14-dimensional fused auxiliary space $V_{\bar 1}$ by the operators
$P^{(14)}_{21}$ or $P^{(14)}_{\tilde 1 2}$, and $\tilde 1$ means the 14-dimensional fused auxiliary space $V_{\tilde 1}$ by the operator
$P^{(14)}_{321}$.

Next, we calculate the products of two monodromy matrices with special spectral parameters. By using the property
of permutation operator, we obtain
\bea &&
T_{a}(\theta_j)T_{b}(\theta_j+\delta)=R_{a1}(\theta_j-\theta_1)\cdots
R_{aj-1}(\theta_j-\theta_{j-1})R_{aj}(0)R_{aj+1}(\theta_j-\theta_{j+1})\cdots \no\\[4pt]
&&\qquad \times R_{aN}(\theta_j-\theta_{N}) R_{b1}(\theta_j-\theta_1+\delta)\cdots
R_{bj-1}(\theta_j-\theta_{j-1}+\delta)R_{bj}(\delta)\no\\[4pt]
&&\qquad \times R_{aj}(0)R_{ja}(0)\rho_v(0)^{-1}R_{bj+1}(\theta_j-\theta_{j+1}+\delta)\cdots
R_{bN}(\theta_j-\theta_{N}+\delta)\no\\[4pt]
&&=R_{jj+1}(\theta_j-\theta_{j+1})\cdots
R_{jN}(\theta_j-\theta_{N}) R_{a1}(\theta_j-\theta_1)\cdots
R_{aj-1}(\theta_j-\theta_{j-1})\no\\[4pt]
&&\qquad \times R_{b1}(\theta_j-\theta_1+\delta)\cdots
R_{bj-1}(\theta_j-\theta_{j-1}+\delta)\no\\[4pt]
&&\qquad \times P_{ba}^{(d)}S_{ba} R_{ja}(0)
R_{bj+1}(\theta_j-\theta_{j+1}+\delta)\cdots
R_{bN}(\theta_j-\theta_{N}+\delta)\no\\[4pt]
&&=P_{ba}^{(d)}  R_{a1}(\theta_j-\theta_1)\cdots
R_{aj-1}(\theta_j-\theta_{j-1})R_{aj}(0)R_{ja}(0)\rho_v(0)^{-1} R_{jj+1}(\theta_j-\theta_{j+1})\cdots \no\\[4pt]
&&\qquad \times R_{jN}(\theta_j-\theta_{N}) R_{b1}(\theta_j-\theta_1+\delta)\cdots
R_{bj-1}(\theta_j-\theta_{j-1}+\delta)\no\\[4pt]
&&\qquad \times R_{ba}(\delta) R_{ja}(0)
R_{bj+1}(\theta_j-\theta_{j+1}+\delta)\cdots
R_{bN}(\theta_j-\theta_{N}+\delta)\no\\[4pt]
&&=P_{ba}^{(d)}T_{a}(\theta_j)T_{b}(\theta_j+\delta),\label{fui-7-aitan-1} \eea
where $\delta$ is the degenerate point of $R_{ab}(u)$ and
$P_{ba}^{(d)}$ is the corresponding $d$-dimensional project operator.
The product of three monodromy matrices at fixed points is
\bea
&& T_{1'}(\theta_j)T_{\langle 2'3'\rangle}(\theta_j-1)=T_{1'}(\theta_j)P_{3'2'}^{(14)}T_{2'}(\theta_j-1)T_{3'}(\theta_j-2)P_{3'2'}^{(14)}\no\\[4pt]
&&=R_{1'1}(\theta_j-\theta_1)\cdots R_{1'j-1}(\theta_j-\theta_{j-1})R_{1'j}(0)R_{1'j+1}(\theta_j-\theta_{j+1})\cdots
R_{1'N}(\theta_j-\theta_{N})\no\\[4pt]
&&\qquad \times R_{2'1}(\theta_j-\theta_1-1)\cdots
R_{2'j-1}(\theta_j-\theta_{j-1}-1)R_{2'j}(-1)R_{2'j+1}(\theta_j-\theta_{j+1}-1)\cdots\no\\[4pt]
&&\qquad \times R_{2'N}(\theta_j-\theta_{N}-1)
R_{3'1}(\theta_j-\theta_1-2)\cdots
R_{3'j-1}(\theta_j-\theta_{j-1}-2)R_{3'j}(-2)\no\\[4pt]
&&\qquad \times [R_{1'j}(0)R_{j1'}(0)\rho_v(0)^{-1} ]R_{3'j+1}(\theta_j-\theta_{j+1}-2)\cdots
R_{3'N}(\theta_j-\theta_{N}-2)P_{3'2'}^{(14)}\no\\[4pt]
&&=R_{jj+1}(\theta_j-\theta_{j+1})\cdots
R_{jN}(\theta_j-\theta_{N})
R_{1'1}(\theta_j-\theta_1)\cdots
R_{1'j-1}(\theta_j-\theta_{j-1})\no\\[4pt]
&&\qquad \times
R_{2'1}(\theta_j-\theta_1-1)\cdots
R_{2'j-1}(\theta_j-\theta_{j-1}-1)R_{3'1}(\theta_j-\theta_1-2)\cdots\no\\[4pt]
&&\qquad \times
R_{3'j-1}(\theta_j-\theta_{j-1}-2)R_{2'1'}(-1)R_{3'1'}(-2) R_{j1'}(0)P_{3'2'}^{(14)}R_{2'j+1}(\theta_j-\theta_{j+1}-1)\cdots\no\\[4pt]
&&\qquad \times R_{2'N}(\theta_j-\theta_{N}-1)R_{3'j+1}(\theta_j-\theta_{j+1}-2)\cdots
R_{3'N}(\theta_j-\theta_{N}-2)P_{3'2'}^{(14)}\no\\[4pt]
&&=R_{jj+1}(\theta_j-\theta_{j+1})\cdots
R_{jN}(\theta_j-\theta_{N})
R_{1'1}(\theta_j-\theta_1)\cdots
R_{1'j-1}(\theta_j-\theta_{j-1})\no\\[4pt]
&&\qquad \times
R_{2'1}(\theta_j-\theta_1-1)R_{2'2}(\theta_j-\theta_2-1)\cdots
R_{2'j-1}(\theta_j-\theta_{j-1}-1)\no\\[4pt]
&&\qquad \times
R_{3'1}(\theta_j-\theta_1-2)R_{3'2}(\theta_j-\theta_2-2)\cdots
R_{3'j-1}(\theta_j-\theta_{j-1}-2)P_{3'2'1'}^{(14)}S_{3'2'1'}S_{3'2'}^{-1}\no\\[4pt]
&&\qquad \times R_{j1'}(0)R_{2'j+1}(\theta_j-\theta_{j+1}-1)\cdots
R_{2'N}(\theta_j-\theta_{N}-1)\no\\[4pt]
 &&\qquad \times R_{3'j+1}(\theta_j-\theta_{j+1}-2)\cdots
R_{3'N}(\theta_j-\theta_{N}-2)P_{3'2'}^{(14)} \no \\[4pt]
&&=P_{3'2'1'}^{(14)}R_{1'1}(\theta_j-\theta_1)\cdots
R_{1'j-1}(\theta_j-\theta_{j-1})R_{1'j}(0)R_{j1'}(0)\rho_v(0)^{-1}\no\\[4pt]
&&\qquad \times
R_{jj+1}(\theta_j-\theta_{j+1})\cdots
R_{jN}(\theta_j-\theta_{N})
R_{2'1}(\theta_j-\theta_1-1)\cdots
R_{2'j-1}(\theta_j-\theta_{j-1}-1)\no\\[4pt]
&&\qquad \times R_{2'1'}(-1) R_{2'j+1}(\theta_j-\theta_{j+1}-1)\cdots
R_{2'N}(\theta_j-\theta_{N}-1)\no\\[4pt]
&&\qquad \times
R_{3'1}(\theta_j-\theta_1-2)\cdots
R_{3'j-1}(\theta_j-\theta_{j-1}-2)R_{3'1'}(-2)R_{j1'}(0)\no\\[4pt]
 &&\qquad \times R_{3'j+1}(\theta_j-\theta_{j+1}-2)\cdots
R_{3'N}(\theta_j-\theta_{N}-2)P_{3'2'}^{(14)}\no\\[4pt]
&&=P_{3'2'1'}^{(14)}T_{1'}(\theta_j)T_{\langle
2'3'\rangle}(\theta_j-1).\label{fui-7-aitan-2}\eea
Substituting $\delta=\{-4, -1, -7/2, -3\}$ into Eq.(\ref{fui-7-aitan-1})
and using the relations (\ref{fu11tt-7}) and (\ref{fui-7-aitan-2}), we obtain
\bea &&T_1(\theta_j)\,T_2(\theta_j-4)=
P^{(1) }_{21}\,T_1(\theta_j)\,T_2(\theta_j-4),  \no \\[6pt]
&&T_1(\theta_j)\,T_2(\theta_j-1)=P_{21}^{(14)}\,T_1(\theta_j)\,T_2(\theta_j-1), \no \\[6pt]
&&T_1(\theta_j)\,T_{\langle23\rangle}(\theta_j-1)=T_1(\theta_j)P_{32}^{(14)}T_2(\theta_j-1)T_3(\theta_j-2)P_{32}^{(14)}
 \no \\[6pt]
&&\qquad =P_{321}^{(14)}T_1(\theta_j)T_{\langle23\rangle}(\theta_j-1),\no \\[6pt]
&&T_2(\theta_j)T_{\bar 1}(\theta_j-\frac{7}{{2}})=P^{(6)}_{\bar{1}2}\,
T_2(\theta_j)T_{\bar 1}(\theta_j-\frac{7}{{2}}), \no \\[6pt]
&&T_2(\theta_j)T_{\tilde 1}(\theta_j-3)= P^{(14)}_{\tilde{1}2}\,T_2(\theta_j)\, T_{\tilde
1}(\theta_j-3). \label{fui-7} \eea
Taking the partial traces of Eq.(\ref{fui-7}) in the
auxiliary spaces and using
the correspondences (\ref{R5-5}) and (\ref{R11-R11}),
we obtain the closed operator product
identities among transfer matrices \bea &&
t^{(p)}(\theta_j)\,t^{(p)}(\theta_j-4)=\prod_{i=1}^N
a(\theta_j-\theta_i)e(\theta_j-\theta_i-4)\times {\rm id},\no  \\
&& t^{(p)}(\theta_j)\,t^{(p)}(\theta_j-1)= \prod_{i=1}^N
\tilde{\rho}_0(\theta_j-\theta_i)\,t_2^{(p)}(\theta_j-\frac{1}{{2}}),\no \\
&& t^{(p)}(\theta_j)\,t_2^{(p)}(\theta_j-\frac{3}{{2}})=
\prod_{i=1}^N\tilde{\rho}_0(\theta_j-\theta_i)
(\theta_j-\theta_i+1)\,t_3^{(p)}(\theta_j-1),\no \\
&&
t^{(p)}(\theta_j)\,t^{(p)}_2(\theta_j-\frac{7}{{2}})=\prod_{i=1}^N
\tilde{\rho}_0(\theta_j-\theta_i)\,t^{(p)}(\theta_j-3),\no  \\
&& t^{ (p)}(\theta_j)t_3^{(p)}(\theta_j-3)=\prod_{i=1}^N
(\theta_j-\theta_i+4)\,t_2^{(p)}(\theta_j-\frac{5}{{2}}), \quad j=1, \cdots N. \label{Op-Product-Periodic-6}
\eea
In the derivation, we have used the property of projector \bea
&&P^{(14)}_{32}P^{(14)}_{321}=P^{(14)}_{32}P^{(14)}_{321}S_{321}S_{321}^{-1}
=P^{(14)}_{32}R_{32}(-1)R_{31}(-2)R_{21}(-1)S_{321}^{-1} \no \\[4pt]
&&\qquad =P^{(14)}_{32}P^{(14)}_{32}S_{32}
R_{31}(-2)R_{21}(-1)S_{321}^{-1}
=P^{(14)}_{32}S_{32} R_{31}(-2)R_{21}(-1)S_{321}^{-1} \no \\[4pt]
&&\qquad =R_{32}(-1)R_{31}(-2)R_{21}(-1)S_{321}^{-1}
=P^{(14)}_{321}S_{321}S_{321}^{-1}=P^{(14)}_{321}. \eea
The asymptotic behaviors of the fused
transfer matrices can be calculated directly
\bea && t^{ (p)}(u)|_{u\rightarrow \pm\infty}= 6u^{2N}\times {\rm  id} +\cdots,\quad  t^{(p)}_2(u)|_{u\rightarrow \pm\infty}=
14u^{2N}\times {\rm  id} +\cdots,\no \\[4pt]
&& t^{(p)}_3(u)|_{u\rightarrow \pm\infty}= 14u^{N}\times {\rm id}
+\cdots.\label{fuwwwtpl-7} \eea

Denote the eigenvalues of the transfer matrices
$t^{(p)}(u)$, $t^{(p)}_2(u)$ and $t^{(p)}_3(u)$ as
$\Lambda^{(p)}(u)$, $\Lambda^{(p)}_2(u)$, and
$\Lambda^{(p)}_3(u)$, respectively. From the operator product
identities (\ref{Op-Product-Periodic-6}), we obtain
the functional relations among the eigenvalues \bea &&
\Lambda^{ (p)}(\theta_j)\,\Lambda^{(p)}(\theta_j-4)=\prod_{i=1}^N
a(\theta_j-\theta_i)\,e(\theta_j-\theta_i-4),\no   \\
&& \Lambda^{ (p)}(\theta_j)\,\Lambda^{ (p)}(\theta_j-1)=
\prod_{i=1}^N
\tilde{\rho}_0(\theta_j-\theta_i)\,\Lambda^{(p)}_2(\theta_j-\frac{1}{{2}}),\no  \\
&&\Lambda^{(p)}(\theta_j)\,\Lambda^{(p)}_2(\theta_j-\frac{3}{{2}})=
\prod_{i=1}^N\tilde{\rho}_0(\theta_j-\theta_i)
(\theta_j-\theta_i+1)\,\Lambda^{(p)}_3(\theta_j-1), \no  \\
&&
\Lambda^{(p)}(\theta_j)\,\Lambda^{(p)}_2(\theta_j-\frac{7}{{2}})=\prod_{i=1}^N
\tilde{\rho}_0(\theta_j-\theta_i)\,\Lambda^{(p)}(\theta_j-3),\no  \\
&&
\Lambda^{(p)}(\theta_j)\,\Lambda^{(p)}_3(\theta_j-3)=\prod_{i=1}^N
(\theta_j-\theta_i+4)\,\Lambda^{(p)}_2(\theta_j-\frac{5}{{2}}), \quad j=1, \cdots N. \label{Eigen-function-relation-6}
\eea
The asymptotic behaviors (\ref{fuwwwtpl-7}) of the fused transfer
matrices lead to
\bea && \Lambda^{ (p)}(u)|_{u\rightarrow \pm\infty}= 6u^{2N} +\cdots,\quad \Lambda^{(p)}_2(u)|_{u\rightarrow \pm\infty}=
14u^{2N}+\cdots,\no \\[4pt]
&& \Lambda^{(p)}_3(u)|_{u\rightarrow \pm\infty}= 14u^{N}
+\cdots.\label{Eigen-function-relation-10} \eea

From the definitions (\ref{1117-1}) and
(\ref{Fused-transfer-matrix-periodic}), we know that the
eigenvalues $\Lambda^{(p)}(u)$ and $\Lambda^{(p)}_2(u)$ are
the polynomials of $u$ with degree $2N$, while $\Lambda^{ (p)}_3(u)$
is a polynomial of $u$ with degree $N$. Hence
the $5N$ functional relations (\ref{Eigen-function-relation-6}) and 3 asymptotic behaviors (\ref{Eigen-function-relation-10})
can completely determine the eigenvalues of $\Lambda^{(p)}(u)$, $\Lambda^{(p)}_2(u)$ and $\Lambda^{(p)}_3(u)$.

\subsection{$T-Q$ relations}

For the simplicity, let us introduce some functions
\bea &&Z^{(p)}_1(u)=A^{(p)}(u)\frac{Q_{p}^{(1)}(u-1)}{Q_{p}^{(1)}(u)},\quad
Z^{(p)}_2(u)=B^{(p)}(u)\frac{Q_{p}^{(1)}(u+1)Q_{p}^{(2)}(u-1)}{Q_{p}^{(1)}(u)Q_{p}^{(2)}(u)},\no\\[4pt]
&&Z^{(p)}_3(u)=B^{(p)}(u)\frac{Q_{p}^{(2)}(u+1)Q_{p}^{(3)}(u-\frac{3}{2})}
{Q_{p}^{(2)}(u)Q_{p}^{(3)}(u+\frac{1}{2})},\quad
Z^{(p)}_4(u)=B^{(p)}(u)\frac{Q_{p}^{(2)}(u+1)Q_{p}^{(3)}(u+\frac{5}{2})}{Q_{p}^{(2)}(u+2)Q_{p}^{(3)}(u+\frac{1}{2})},\no\\[4pt]
&&Z^{(p)}_5(u)=B^{(p)}(u)\frac{Q_{p}^{(1)}(u+{2})Q_{p}^{(2)}(u+3)}{Q_{p}^{(1)}(u+3)Q_{p}^{(2)}(u+2)},\quad
Z^{(p)}_6(u)=V^{(p)}(u) \frac{Q_{p}^{(1)}(u+4)}{Q_{p}^{(1)}(u+3)},
\eea
where
\bea && A^{(p)}(u)=\prod_{j=1}^N a(u-\theta_j), \quad B^{(p)}(u)=\prod_{j=1}^N b(u-\theta_j), \quad V^{(p)}(u)=\prod_{j=1}^N e(u-\theta_j),\no\\
&&Q_{p}^{(m)}(u)=\prod_{k=1}^{L_m}(u-\mu_k^{(m)}+\frac{m}{2}),
\quad m=1,2,3.
\eea
By using above functions, the eigenvalues of transfer matrices can be expressed as the $T-Q$ relations
\bea
&&\Lambda^{(p)}(u)=\sum_{l=1}^6 Z^{(p)}_l(u), \no \\[4pt]
&&\Lambda^{(p)}_2(u)=\prod_{i=1}^N
\tilde{\rho}_0^{-1}(u-\theta_i+\frac12)\left\{\sum_{i<j}^6 Z^{(p)}_i(u+\frac12) Z^{(p)}_j(u-\frac12)-Z^{(p)}_3(u+\frac12) Z^{(p)}_3(u-\frac12)\right\}
\no  \\[4pt]
&&\Lambda^{(p)}_3(u)=\prod_{i=1}^N [\tilde{\rho}_0(u-\theta_i+1)
\tilde{\rho}_0(u-\theta_i)(u-\theta_i+2)]^{-1}\no\\
&&\hspace{5mm}\times \left\{\sum^6_{i<j<k}
Z^{(p)}_i(u+1)Z^{(p)}_j(u)Z^{(p)}_k(u-1)-\sum_{k=5}^6 Z^{(p)}_3(u+1)Z^{(p)}_4(u)Z^{(p)}_k(u-1) \right. \no\\
&&\hspace{5mm}\left.-\sum_{i=1}^2 Z^{(p)}_i(u+1)Z^{(p)}_3(u)Z^{(p)}_4(u-1)-\sum_{j=3}^4 Z^{(p)}_2(u+1)Z^{(p)}_j(u)Z^{(p)}_5(u-1)\right\}.
\label{T-Q-Hom-3}\eea
All the eigenvalues are polynomials, thus the residues of right hand sides of Eq.(\ref{T-Q-Hom-3}) should be zero, which gives that
the Bethe roots $\{\mu^{(m)}_k\}$ in (\ref{T-Q-Hom-3}) should satisfy the
Bethe ansatz equations \bea &&
\frac{Q_{p}^{(1)}(\mu_k^{(1)}+\frac{1}{2})Q_{p}^{(2)}(\mu_k^{(1)}-\frac{3}{2})}{Q_{p}^{(1)}(\mu_k^{(1)}-\frac{3}{2})Q_{p}^{(2)}(\mu_k^{(1)}-\frac{1}{2})}
=-\prod_{j=1}^N \frac{\mu_k^{(1)}+\frac{1}{2}-\theta_j
}{\mu_k^{(1)}-\frac{1}{2}-\theta_j}, \quad k=1,\cdots, L_1, \no \\[8pt]
&&\frac{Q_{p}^{(1)}(\mu_l^{(2)})Q_{p}^{(2)}(\mu_l^{(2)}-2)Q_{p}^{(3)}(\mu_l^{(2)}-\frac{1}{2})}
{Q_{p}^{(1)}(\mu_l^{(2)}-1)Q_{p}^{(2)}(\mu_l^{(2)})Q_{p}^{(3)}(\mu_l^{(2)}-\frac{5}{2})}
=-1, \quad l=1,\cdots, L_2,\no\\[8pt]
&&\frac{Q_{p}^{(2)}(\mu_l^{(3)}+\frac{1}{2})Q_{p}^{(3)}(\mu_l^{(2)}-3)}
{Q_{p}^{(2)}(\mu_l^{(3)}-\frac{1}{2})Q_{p}^{(3)}(\mu_l^{(3)}+1)}
=-1, \quad l=1,\cdots, L_3.\label{BAEs-3} \eea
We note that the Bethe ansatz equations obtained from the regularities of
$\Lambda^{(p)}(u)$ are the same as those obtained from $\Lambda^{(p)}_2(u)$ and $\Lambda^{(p)}_3(u)$. Meanwhile, any of these eigenvalues can give the complete set of
Bethe ansatz equations.

It is easy to check that the $T-Q$ relations (\ref{T-Q-Hom-3})
satisfy the functional relations (\ref{Eigen-function-relation-6}) and the asymptotic behaviors (\ref{Eigen-function-relation-10}).
Therefore, we conclude that the $\Lambda^{(p)}(u)$, $\Lambda^{(p)}_2(u)$ and $\Lambda^{(p)}_3(u)$ are the eigenvalues of
the transfer matrices $t^{(p)}(u)$, $t^{(p)}_2(u)$ and $t^{(p)}_3(u)$, respectively, provided that the Bethe roots
satisfy the Bethe ansatz equations (\ref{BAEs-3}).
It is remarked that the $T-Q$ relations (\ref{T-Q-Hom-3}) and the associated Bethe ansatz equations
(\ref{BAEs-3}) (after taking the homogeneous limit
$\{\theta_j\to 0|j=1,2,\cdots,N\}$) coincide with the previous results \cite{NYRes,NYReshetikhin2,Bn}.
Then the eigenvalues of the Hamiltonian (\ref{asdasd}) reads
\begin{eqnarray}
E_p= \frac{\partial \ln \Lambda^{(p)}(u)}{\partial
u}|_{u=0,\{\theta_j\}=0}.
\end{eqnarray}

\section{$C_3$ model with open boundary condition}
\setcounter{equation}{0}

\subsection{Boundary integrability}

Now, we consider the system with open boundary condition.
The boundary reflections are quantified by the reflection matrix $K^-$ at one side and dual one $K^+$ at the other side.
The integrable requires that $K^-$ satisfies the reflection equation
\begin{equation}
 R_{12}(u-v){K^{-}_{  1}}(u)R_{21}(u+v) {K^{-}_{2}}(v)=
 {K^{-}_{2}}(v)R_{12}(u+v){K^{-}_{1}}(u)R_{21}(u-v),
 \label{r1}
 \end{equation}
while $K^+$ satisfies the dual reflection equation
\begin{eqnarray}
 &&R_{12}(-u+v){K^{+}_{1}}(u)R_{21}
 (-u-v-8){K^{+}_{2}}(v)\nonumber\\[4pt]
&&\qquad\qquad\quad\quad={K^{+}_{2}}(v)R_{12}(-u-v-8)
{K^{+}_{1}}(u)R_{21}(-u+v).
 \label{r2}
 \end{eqnarray}
In this paper, we consider the case that the reflection matrices have
off-diagonal elements, thus the numbers of quasi-particles with different intrinsic degrees of freedom are not conserved during the reflection processes.
The reflection matrix $K_0^{-}(u)$ defined in the space $V_0$ takes the form of \cite{8Mel054,8Mel055,8Mel056}
\bea
K_0^{-}(u)=\zeta+M u, \quad
M=\left(\begin{array}{cccccc}-1&0 &0&c_{1}&0&0\\[6pt]
    0&-1 &0&0&c_{1}&0\\[6pt]
    0&0&-1&0&0 &c_{1} \\[6pt]
    c_{2}&0&0&1&0&0\\[6pt]
   0& c_{2}&0&0&1&0\\[6pt]
  0& 0& c_{2}&0&0&1\\[6pt]
 \end{array}\right),\label{K-matrix-VV}\eea
where $\zeta$, $c_1$ and $c_2 $ are the arbitrary boundary parameters.
The dual reflection matrix $K_0^{+}(u)$ is defined as
\begin{equation}
K_0^{ +}(u)=K_0^{ -}(-u-4)|_{\zeta,c_i\rightarrow
    \tilde{\zeta},\tilde{c}_i }, \label{ksk}
\end{equation}
where $\tilde{\zeta}$ and $\tilde{c}_i (i=1, 2)$ are the
boundary parameters.

Due to the boundary reflection, besides the monodromy matrix $T_0(u)$ given by (\ref{Mon-1}), we also need the reflecting monodromy matrix
\begin{eqnarray}
\hat{T}_0 (u)=R_{N0}(u+\theta_N)\cdots R_{20}(u+\theta_{2}) R_{10}(u+\theta_1),\label{Tt11}
\end{eqnarray}
which satisfies the Yang-Baxter relation
\begin{eqnarray}
R_{ 12} (u-v) \hat T_{1}(u) \hat T_2(v)=\hat  T_2(v) \hat T_{ 1}(u) R_{12} (u-v).\label{haishi0}
\end{eqnarray}
The transfer matrix $t(u)$ of the system with open boundary condition is
\begin{equation}
t(u)= tr_0 \{K_0^{+}(u)T_0 (u) K^{-}_0(u)\hat{T}_0 (u)\}. \label{trweweu}
\end{equation}

From the Yang-Baxter relations (\ref{ybta2o}) and (\ref{haishi0}), reflection equation (\ref{r1}) and dual one (\ref{r2}), we can
prove that the transfer matrices (\ref{trweweu}) with different spectral parameters
commutate with each other, i.e., $[t(u), t(v)]=0$. Therefore, $t(u)$ serves
as the generating function of all the conserved quantities of the
system. The Hamiltonian is constructed as
the derivative of logarithm of the transfer matrix
\begin{eqnarray}
H=\frac{\partial \ln t(u)}{\partial
u}|_{u=0,\{\theta_j\}=0}. \label{hh}
\end{eqnarray}
In the Hamiltonian (\ref{hh}), because two boundary reflection matrices $K_0^{-}(u)$ (\ref{K-matrix-VV}) and $K_0^{+}(u)$ (\ref{ksk}) do not commutate
with each other, i.e., $[K_0^-(u),\,K_0^+(v)]\neq 0$, the $K_0^{\pm}(u)$ cannot be
diagonalized simultaneously. Then it is quite hard to derive the exact solutions of the system via the conventional Bethe Ansatz due to the
absence of a proper reference state. We will generalize the method
developed in section 2 to calculate the eigenvalues of transfer matrix
(\ref{trweweu}) and that of Hamiltonian (\ref{hh}) in the following subsections.

\subsection{Fusion}

Because the reflection matrices are defined in the auxiliary spaces and we have fused the auxiliary spaces
into different forms with different dimensions, we should fuse the reflection matrices correspondingly.
All the fusion relations with boundary reflections can be obtain from the degeneration properties of $R$-matrix
and the (dual) reflection equation. The related projectors are
$P^{(1)}_{12}$, $P_{12}^{(14)}$,
$P_{123}^{(14)}$, $P^{(6)}_{\bar{1}2}$ and $P^{(14)}_{\tilde{1}2}$ defined above.
The fusion of reflection matrices with one-dimensional projector gives
\bea && P_{21}^{(1)}K_{1}^{-}(u)R_{21}(2u-4)
K_{2}^{-}(u-4)P_{12}^{(1)}={\rm Det}_q(K^{-}(u))\,P_{12}^{ (1)},\no \\
&& P_{12}^{(1)}K_{2}^{+}(u-4)R_{12}(-2u-4)K_{1}^{
+}(u)P_{21}^{(1)}= {\rm
Det}_q(K^{+}(u))\,P_{21}^{(1)}, \label{fk-2}\eea where ${\rm Det}_q(K^{\pm}(u))$ are the quantum determinants of reflection matrices $K^{\pm}(u)$,
\bea
&&{\rm Det}_q(K^{-}(u))=(u-\frac{3}{{2}})(u-4)h_1(u)h_2(u),
\no\\
&&{\rm Det}_q(K^{+}(u))
=(u+\frac{3}{{2}})(u+4)\tilde{h}_1(u)\tilde{h}_2(u),\no\\
&& h_1(u)=2(\sqrt{(1+c_1c_2)}u+\zeta),\quad
h_2(u)=2(\sqrt{(1+c_1c_2)}u-\zeta),\no\\
&&\tilde{h}_1(u)=-2(\sqrt{1+\tilde{c}_1\tilde{c}_2}u-\tilde{\zeta}),\quad
\tilde{h}_2(u)=-2(\sqrt{1+\tilde{c}_1\tilde{c}_2}u+\tilde{\zeta}).
\eea We note that the reflection equation and dual one require
that the inserted $R$-matrices in (\ref{fk-2}) with determined
spectral parameters are necessary.

Using the $14$-dimensional projector $P_{12}^{(14)}$, we construct the $14\times 14$ fused
$K$-matrices \bea &&
K_{\langle12\rangle}^-(u+\frac{1}{2})
=\frac{1}{2(u-\frac{1}{2})(u+2)}
P_{21}^{(14)}K_{1}^{  -}(u+\frac{1}{2})R_{21}(2u)K_{2}^{ -}(u-\frac{1}{2})P_{12}^{(14)}\equiv  K^{
-}_{\bar{1}}(u),\no\\
&&
K_{\langle12\rangle}^+(u+\frac{1}{2})=
\frac{1}{2(u+2)(u+\frac{9}{2})}P_{12}^{(14)}K_{2}^{
+}(u-\frac{1}{2})
R_{12}(-2u-8)K_{1}^{
+}(u+\frac{1}{2})P_{21}^{(14)} \no\\[4pt]
&&\qquad \qquad  \;\;\quad \equiv K^{ +}_{\bar{1}}(u).\label{K-matrix-Bar+}
\eea
The fused reflection matrices (\ref{K-matrix-Bar+}) satisfy the reflection equations
\begin{eqnarray}
&& R_{\bar 12}(u-v) K^{-}_{\bar  1}(u) R_{2\bar 1}(u+v) K^{-}_{2}(v)=K^{-}_{2}(v)
R_{\bar 12}(u+v) K^{-}_{\bar 1}(u) R_{2\bar 1}(u-v), \no \\[4pt]
&& R_{\bar 12}(-u+v)  K^{+}_{\bar 1}(u) R_{2\bar 1} (-u-v-8) K^{ +}_{2}(v) \nonumber\\[4pt]
&&\qquad =K^{ +}_{2}(v) R _{\bar 12}(-u-v-8) K^{ +}_{\bar 1}(u) R _{2\bar 1}(-u+v), \label{haaishi8}
\end{eqnarray}
which means that the fusion does not break the integrability.

The 14-dimensional projector $P_{123}^{(14)}$ allows us to construct the $14\times 14$
fused $K$-matrices \bea
&&
K_{\langle123\rangle}^-(u+1)
=[2^3(u+\frac{5}{2})(u+\frac{3}{2})
(u-\frac{1}{2})u(u-1)(u+2)]^{-1}\nonumber\\[4pt]
&&\;\;\times P_{321}^{(14)}K_{1}^{
-}(u+1)R_{21}(2u+1)R_{31}(2u) K^{  -}_{2}(u)R_{32}(2u-1)K_{3}^{  -}(u-1)P_{123}^{(14)}\equiv  K^{-}_{\tilde{1}}(u),\no \\[4pt]
&&
K_{\langle123\rangle}^+(u+1)
=[2^3(u+\frac{3}{2})(u+\frac{5}{2})
(u+\frac{9}{2})(u+2)(u+4)(u+5)]^{-1}P_{123}^{(14)}K_{3}^{ +}(u-1)\nonumber\\[4pt]
&&\;\;\times R_{23}(-2u-7)R_{13}(-2u-8)K^{+}_{2}(u)R_{12}(-2u-9) K_{1}^{
+}(u+1)P_{321}^{(14)}\equiv K^{+}_{\tilde{1}}(u).\label{K-matrix-tilde+} \eea
The fused reflection matrix (\ref{K-matrix-tilde+}) satisfy the reflection equations
\begin{eqnarray}
&& R_{\tilde 12}(u-v) K^{-}_{\tilde  1}(u) R_{2\tilde 1}(u+v) K^{-}_{2}(v)=K^{-}_{2}(v)
R_{\tilde 12}(u+v) K^{-}_{\tilde 1}(u) R_{2 \tilde 1}(u-v), \no \\[4pt]
&& R_{\tilde 12}(-u+v)  K^{+}_{\tilde 1}(u) R_{2\tilde 1} (-u-v-8) K^{ +}_{2}(v) \nonumber\\[4pt]
&&\qquad =K^{ +}_{2}(v) R _{\tilde 12}(-u-v-8) K^{ +}_{\tilde 1}(u) R _{2\tilde 1}(-u+v), \no\\[4pt]
&& R_{\tilde 1\bar 2}(u-v) K^{-}_{\tilde  1}(u) R_{\bar 2\tilde 1}(u+v) K^{-}_{\bar 2}(v)=K^{-}_{\bar 2}(v)
R_{\tilde 1\bar 2}(u+v) K^{-}_{\tilde 1}(u) R_{\bar 2 \tilde 1}(u-v), \no \\[4pt]
&& R_{\tilde 1\bar 2}(-u+v)  K^{+}_{\tilde 1}(u) R_{\bar 2\tilde 1} (-u-v-8) K^{ +}_{\bar 2}(v) \nonumber\\[4pt]
&&\qquad =K^{ +}_{\bar 2}(v) R _{\tilde 1\bar 2}(-u-v-8) K^{ +}_{\tilde 1}(u) R _{\bar 2\tilde 1}(-u+v). \label{haaishdfi811-1}
\end{eqnarray}

Using the $6$-dimensional projectors $P_{\bar{1}2}^{(6)}$ and the
correspondence (\ref{Identification-1}), we have \bea
&&K^{-}_{\langle
\bar{1}2\rangle }(u+3)= \frac{P_{\bar{1}2}^{(6)}K_{2}^{
-}(u+3) R_{\bar 12}(2u+\frac{5}2)K_{\bar{1}}^{-}(u-\frac{1}{2})P_{2\bar{1}}^{ (6)}}
{2(u+2)(u-\frac{1}{2})h_1(u+3)h_2(u+3)}\equiv  K^{-}_1(u), \no \\[4pt]
&&K^{+}_{\langle
\bar{1}2\rangle }(u+3)= \frac{P_{2\bar{1}}^{(6)}K_{
\bar{1}}^{+}(u-\frac{1}{2})R_{2\bar{1}}(-2u-\frac{21}{2})K_{2}^{+}(u+3)P_{\bar{1}2}^{(6)}}
{2(u+7)(u+\frac{9}{2})\tilde{h}_1(u+3)\tilde{h}_2(u+3)}\equiv K^{+}_1(u).\label{haaishi811-1} \eea
We note the fused reflection matrices (\ref{haaishi811-1}) are the same as the original ones given by (\ref{K-matrix-VV}) and (\ref{ksk}).
Similarly, with the help of the $14$-dimensional projectors
$P_{\tilde{1}2}^{(14)}$ and the correspondence
(\ref{Identification-2}), we have \bea
&&K^{
-}_{\langle \tilde{1}2\rangle }(u+\frac{5}{2})=-
\frac{P_{\tilde{1}2}^{ (14)}K_{2}^{
-}(u+\frac{5}{2})R_{\tilde{1}2}(2u+2)
K_{\tilde{1}}^{ -}(u-\frac{1}{2})P_{2\tilde{1}}^{
(14)}}
{2(u-\frac{1}{2})h_1(u+\frac{5}{2})h_2(u+\frac{5}{2})}\equiv K^{-}_{\bar{1}}(u),\no \\[4pt]
&&K^{+}_{\langle \tilde{1}2\rangle }(u+\frac{5}{2})=
\frac{P_{2\tilde{1}}^{(14)}K_{\tilde{1}}^{
+}(u-\frac{1}{2})R_{2\tilde{1}}(-2u-10)
K_{2}^{+}(u+\frac{5}{2})P_{\tilde{1}2}^{(14)}}
{2(u+\frac{13}{2})\tilde{h}_1(u+\frac{5}{2})\tilde{h}_2(u+\frac{5}{2})}\equiv K^{+}_{\bar{1}}(u).\label{K-bar+}\eea
We note that the fused reflection matrices (\ref{K-bar+}) are the same as the fused ones (\ref{K-matrix-Bar+}).
Now we have obtain all the necessary fused reflection matrices, which are used to construct the conserved quantities and fusion relations of the system
with open boundary conditions.

\subsection{Operator product identities}

The fused reflecting monodromy matrices are defined as
\begin{eqnarray}
&&\hat{T}_{\bar{0}}(u)=R_{N\bar{0}}(u+\theta_N)\cdots R_{2\bar{0}}(u+\theta_{2}) R_{1\bar{0}}(u+\theta_1),\no \\[4pt]
 &&\hat{T}_{\tilde {0}}(u)=R_{N\tilde {0}}(u+\theta_N)\cdots R_{2\tilde {0}}(u+\theta_{2}) R_{1\tilde {0}}(u+\theta_1).\label{Mon-2}
\end{eqnarray}
where $R_{2\bar{1}}(u)$ and $R_{2\tilde{1}}(u)$ can be obtained from the first relations in (\ref{lic-13}) and (\ref{lic-23}), respectively.
The fused reflecting monodromy matrices satisfy the Yang-Baxter relations
\begin{eqnarray}
&&R_{\bar 12} (u-v) \hat{T}_{\bar{1}}(u)\hat{T}_{2}(v)=\hat{T}_{2}(v)\hat{T}_{\bar{1}}(u)R_{\bar 12} (u-v), \no  \\[4pt]
&&R_{\tilde  12} (u-v) \hat{T}_{\tilde 1}(u)\hat{T}_{2}(v)=\hat{T}_{2}(v)\hat{T}_{\tilde 1}(u)R_{\tilde  12} (u-v), \no  \\[4pt]
&&R_{\bar 1\tilde  1} (u-v) \hat{T}_{\bar 1}(u)\hat{T}_{\tilde 2}(v)=\hat{T}_{\tilde 2}(v)\hat{T}_{\bar 1}(u)R_{\bar 1\tilde 2} (u-v).\label{M2on-2}
\end{eqnarray}
The fused transfer matrices are the partial traces of fused monodromy matrices
\bea
&&t_2(u)=tr_{\bar{0}}\{K^{+}_{\bar{0}}(u) T_{\bar{0}}(u)K^{-}_{\bar{0}}(u)\hat{T}_{\bar{0}}(u)\},\no\\[4pt]
&&t_3(u)=tr_{\tilde{0}}\{K^{+}_{\tilde{0}}(u)
T_{\tilde{0}}(u)K^{-}_{\tilde{0}}(u)\hat{T}_{\tilde{0}}(u)\},\label{2c0-1}
\eea
where the fused reflection matrices $K_{\bar{0}}^{\pm}(u)$ and $K_{\tilde 0}^{\pm}(u)$ are given by
(\ref{K-matrix-Bar+}) and
(\ref{K-matrix-tilde+}), respectively. From the Yang-Baxter relations (\ref{ybta2o-2}),
(\ref{M2on-2}) and reflection equations (\ref{haaishi8}), (\ref{haaishdfi811-1}), one can prove that the transfer matrices $t(u)$, $t_2(u)$ and $t_3(u)$
commutate with each other
\bea
[t(u),\,t_2(u)]=[t(u),\,t_3(u)]=[t_2(u),\,t_3(u)]=0.
\eea
Thus these transfer matrices have common eigenstates and can be diagonalized simultaneously.

In order to solve these transfer matrices, we should seek the constraints they satisfied.
The method is fusion. The fusions of reflecting monodromy matrices read
\bea &&P^{(1) }_{12}\hat{T}_1(u)\hat{T}_2(u-4)P^{(1) }_{12}=\prod_{i=1}^N
a(u+\theta_i)e(u+\theta_i-4)P^{(1)}_{12}\times {\rm id},  \no \\[4pt]
&&P_{12}^{(14)}\hat{T}_1(u)\hat{T}_2(u-1)P_{12}^{(14)}=\hat{T}_{\langle12\rangle}(u)=\prod_{i=1}^N
\tilde{\rho}_0(u+\theta_i)\hat{T}_{\bar{1}}(u-\frac{1}{{2}}),\no \\[4pt]
&&P_{123}^{(14)}\hat{T}_1(u)\hat{T}_2(u-1)\hat{T}_3(u-2)P_{123}^{(14)}\no\\[4pt]
&&\hspace{20mm} =\prod_{i=1}^N
\tilde{\rho}_0(u+\theta_i)\tilde{\rho}_0(u+\theta_i-1)(u+\theta_i+1)\hat{T}_{\tilde{1}}(u-1),\no\\[4pt]
&&P^{(6) }_{2\bar{1}}\hat{T}_2(u)\hat{T}_{\bar{1}}(u-\frac{7}{{2}})P^{(6) }_{2\bar{1}}=\prod_{i=1}^N
\tilde{\rho}_0(u+\theta_i)\hat{T}_{1}(u-3),\no\\[4pt]
&&P^{(14) }_{2\tilde{1}}\hat{T}_2(u)\hat{T}_{\tilde{1}}(u-3)P^{(14)
}_{2\tilde{1}}=\prod_{i=1}^N (u+\theta_i+4)\hat{T}_{\bar 1}(u-\frac{5}{{2}}).\label{futt-7}
\eea
Meanwhile, the products of reflecting monodromy matrices at two special points satisfy
\bea &&\hat{T}_1(-\theta_j)\hat{T}_2(-\theta_j-4)=P^{(1)}_{12}\hat{T}_1(-\theta_j)\hat{T}_2(-\theta_j-4),\no\\
&&\hat{T}_1(-\theta_j)\hat{T}_2(-\theta_j-1)=P_{12}^{(14)}\hat{T}_1(-\theta_j)\hat{T}_2(-\theta_j-1),\no\\[4pt]
&&\hat{T}_1(-\theta_j)\hat{T}_{\langle23\rangle}(-\theta_j-1)=P_{123}^{(14)}\hat{T}_1(-\theta_j)\hat{T}_{\langle23\rangle}(-\theta_j-1),\no\\[4pt]
&&\hat{T}_2(-\theta_j)\hat{T}_{\bar 1}(-\theta_j-\frac{7}{{2}})=P^{(14)}_{2\bar{1}}\hat{T}_2(-\theta_j)\hat{T}_{\bar 1}(-\theta_j-\frac{7}{{2}}),\no\\[4pt]
&&\hat{T}_2(-\theta_j)\hat{T}_{\tilde 1}(-\theta_j-3)=P^{(14)}_{2\tilde{1}}\hat{T}_2(-\theta_j)\hat{T}_{\tilde 1}(-\theta_j-3).\label{fuii-7} \eea

Then, we are read to consider the constraints of transfer matrices. Direct calculation shows
\bea && t_a(u)t_b(u+\delta)=tr_a\{K_a^{ +}(u)T_a (u) K^{
-}_a(u)\hat{T}_a
(u)\}\no\\[4pt]
&&\hspace{10mm}\times tr_{b}\{K^{+}_{b}(u+\delta)
T_{b}(u+\delta)K^{
-}_{b}(u+\delta)\hat{T}_{b}(u+\delta)\}^{t_b}\no\\[4pt]
&&\hspace{7mm}=tr_{ab}\{K_a^{+}(u)T_a (u) K^{ -}_a(u)\hat{T}_a
(u) [T_{b}(u+\delta)K^{
-}_{b}(u+\delta)\hat{T}_{b}(u+\delta)]^{t_b}[K^{+}_{b}(u+\delta)]^{t_b}\}\no\\[4pt]
&&\hspace{7mm}=[\tilde{\rho}_{ab}(2u+\delta)]^{-1}tr_{ab} \{K_a^{
+}(u)T_a (u) K^{ -}_a(u)\hat{T}_a(u) [T_{b}(u+\delta)K^{
-}_{b}(u+\delta)\no\\[4pt]
&&\hspace{10mm}\times  \hat{T}_{b}(u+\delta)]^{t_b}
R_{ba}^{t_b}(2u+\delta)R_{ab}^{t_b}(-2u-8-\delta)
[K^{+}_{b}(u+\delta)]^{t_b}\}\no\\[4pt]
&&\hspace{7mm}=[\tilde{\rho}_{ab}(2u+\delta)]^{-1}tr_{ab}\{[K^{+}_{b}(u+\delta)
R_{ab}(-2u-8-\delta) K_a^{ +}(u)T_a (u) \no\\[4pt]
&&\hspace{10mm}\times  K^{ -}_a(u)\hat{T}_a
(u)]^{t_b}[R_{ba}(2u+\delta)T_{b}(u+\delta)K^{
-}_{b}(u+\delta)\hat{T}_{b}(u+\delta)]^{t_b} \}\no\\[4pt]
&&\hspace{7mm}=[\tilde{\rho}_{ab}(2u+\delta)]^{-1}tr_{ab}\{K^{+}_{b}(u+\delta)
R_{ab}(-2u-8-\delta)K_a^{ +}(u)T_a (u) \no\\[4pt]
&&\hspace{10mm}\times  K^{ -}_a(u)\hat{T}_a
(u)R_{ba}(2u+\delta)T_{b}(u+\delta)K^{
-}_{b}(u+\delta)\hat{T}_{b}(u+\delta) \}\no\\[4pt]
&&\hspace{7mm}=[\tilde{\rho}_{ab}(2u+\delta)]^{-1}tr_{ab}\{K^{+}_{b}(u+\delta)
R_{ab}(-2u-8-\delta) K_a^{ +}(u)T_a (u) T_{b}(u+\delta)\no\\
&&\hspace{10mm}\times K^{ -}_a(u)R_{ba}(2u+\delta)K^{
-}_{b}(u+\delta) \hat{T}_a(u)\hat{T}_{b}(u+\delta)\}.\label{t-tsd1-1}\eea
In the derivation, we
have used the relations \bea
&&tr_{ab}\{A_{ab}^{t_a}B_{ab}^{t_a}\}=tr_{ab}\{A_{ab}^{t_b}B_{ab}^{t_b}\}
=tr_{ab}\{A_{ab}B_{ab}\},\no \\[4pt]
&& \hat{T}_a (u)  R_{ba}(2u+\delta)T_{b}(u+\delta)=
T_{b}(u+\delta)R_{ba}(2u+\delta)\hat{T}_a (u), \no \\[4pt]
 && R_{ba}^{t_b}(2u+\delta)R_{ab}^{t_b}(-2u-8-\delta)=\tilde{\rho}_{ab}(2u+\delta).\no \eea
The following relation also holds \bea
&&t_1(u)t_{\langle 23 \rangle }(u-1)=t_1(u)tr_{23}\{P_{23}^{(14)}K_3^+(u-2)R_{23}(-2u-5)K_2^+(u-1)
\no\\[4pt]
&&\quad \times
T_2(u-1)T_3(u-2)K_2^-(u-1)R_{32}(2u-3)K_3^-(u-2)\hat{T}_2(u-1)\hat{T}_3(u-2)P_{23}^{(14)}\}\no\\[4pt]
&&=\tilde{\rho}_v(2u-1)\tilde{\rho}_v(2u-2)tr_{123}
\{K^+_3(u-2)R_{23}(-2u-5)K^+_2(u-1)R_{13}(-2u-6)\no\\[4pt]
&&\quad \times R_{12}(-2u-7)K^+_1(u)T_1(u)T_{\langle 23 \rangle }(u-1)K_1^-(u)R_{21}(2u-1)R_{31}(2u-2)\no\\[4pt]
&&\quad \times
K_2^-(u-1)R_{32}(2u-3)K_3^-(u-2)\hat{T}_1(u)\hat{T}_{\langle 23 \rangle }(u-1)\}.\label{t-tsd1-2}\eea

Substituting $u=\pm\theta_j$, $\delta=\{-4, -1, -7/2, -3\}$ into Eqs.(\ref{t-tsd1-1}) and (\ref{t-tsd1-2}),
and using the relations (\ref{fu11tt-7}), (\ref{fui-7}), (\ref{futt-7}), (\ref{fuii-7}) and
the forms of reflection matrices,
%(\ref{K-matrix-VV}), (\ref{ksk}), (\ref{K-matrix-Bar+}), (\ref{K-matrix-tilde+}),
we obtain the closed operator product identities among fused transfer matrices
\bea && t(\pm\theta_j)t(\pm\theta_j-4)={\frac{1}{2^4}}
\frac{(\pm\theta_j-\frac{3}{{2}})
(\pm\theta_j+\frac{3}{{2}})(\pm\theta_j-4)(\pm\theta_j+4)}
{(\pm\theta_j-\frac{1}{{2}})(\pm\theta_j+\frac{1}{{2}})
(\pm\theta_j-2)(\pm\theta_j+2)}\no\\[4pt]
&&\hspace{20mm}\times H_1(\pm\theta_j)H_2(\pm\theta_j)\varrho(\pm\theta_j)\varrho(\mp\theta_j), \no \\[4pt]
&& t(\pm\theta_j)t(\pm\theta_j-1)= \frac{1}{ 2^{2}} \frac{
(\pm\theta_j-1)(\pm\theta_j+\frac{3}{{2}})^2(\pm\theta_j+4)}
{(\pm\theta_j-\frac{1}{{2}})(\pm\theta_j+\frac{7}{{2}})
(\pm\theta_j+1)(\pm\theta_j+2)}\nonumber\\[4pt]
&&\hspace{20mm}\times \varrho(\pm\theta_j)t_2(\pm\theta_j-\frac{1}{{2}}), \no \\[4pt]
&& t(\pm\theta_j)t_2 (\pm\theta_j-\frac{3}{{2}})=\frac{1}{2^4}
\frac{(\pm\theta_j-\frac{3}{2})(\pm\theta_j+\frac{3}{{2}})
(\pm\theta_j+1)(\pm\theta_j+4)}{(\pm\theta_j-\frac{1}{{2}})(\pm\theta_j+\frac{1}{{2}})
(\pm\theta_j+2)(\pm\theta_j+3)}\nonumber\\[4pt]
&&\hspace{20mm}\times
\varrho(\pm\theta_j)\prod_{i=1}^N(\pm\theta_j-\theta_i+1)(\pm\theta_j+\theta_i+1)
t_3(\pm\theta_j-1), \no \\[4pt]
&& t(\pm\theta_j)t_2(\pm\theta_j-\frac{7}{2})=\frac{1}{ 2^{2}}
\frac{(\pm\theta_j-1)(\pm\theta_j-\frac{7}{{2}})(\pm\theta_j+\frac{3}{{2}})
(\pm\theta_j+4)}{(\pm\theta_j-\frac{1}{{2}})(\pm\theta_j-\frac{3}{{2}})(\pm\theta_j+1)
(\pm\theta_j+2)}\nonumber\\[4pt]
&&\hspace{20mm}\times
 H_1(\pm\theta_j)H_2(\pm\theta_j)\varrho(\pm\theta_j)t(\pm\theta_j-3),\no \\[4pt]
&& t(\pm\theta_j)t_3(\pm\theta_j-3)=  \frac{
(\pm\theta_j-3)(\pm\theta_j+4)}{(\pm\theta_j-1)(\pm\theta_j+2) }\prod_{i=1}^N(\pm\theta_j-\theta_i+4)(\pm\theta_j+\theta_i+4)
\no\\
&&\hspace{20mm}\times H_1(\pm\theta_j)H_2(\pm\theta_j) t_2(\pm\theta_j-\frac{5}{2}),\quad j=1, \cdots, N,
\label{futpp-7} \eea where \bea
H_1(u)=h_1(u)\tilde{h}_1(u), \quad H_2(u)=h_2(u)\tilde{h}_2(u), \quad \varrho(u)=\prod_{i=1}^N\tilde{\rho}_0(u-\theta_i)\tilde{\rho}_0(u+\theta_i).
\no\eea
The asymptotic of transfer matrices can be derived directly
\bea && t(u)|_{u\rightarrow \pm\infty}=
-3(2+c_1\tilde{c}_2+c_2\tilde{c}_1)
 u^{4N+2}\times {\rm id}+\cdots, \no \\[4pt]
&& t_2(u)|_{u\rightarrow \pm\infty}=  2^{2}[3(2+c_1\tilde{c}_2+c_2\tilde{c}_1)^2+2(1+{c}_1{c}_2)
(1+\tilde{c}_1\tilde{c}_2)] u^{4N+4}\times{\rm id}+\cdots,\no \\[4pt]
&&t_3(u)|_{u\rightarrow \pm\infty}=-
2^{6}(2+c_1\tilde{c}_2+c_2\tilde{c}_1)[(2+c_1\tilde{c}_2+c_2\tilde{c}_1)^2+3(1+{c}_1{c}_2)(1+\tilde{c}_1\tilde{c}_2)]
u^{2N+6}\times {\rm id}\no\\&&\hspace{25mm}+\cdots. \label{as-tv2}
\eea
According to the definitions, we also know \bea
&&t(0)=6\zeta\tilde{\zeta} \prod_{l=1}^N\rho_1
(-\theta_l)\,\times{\rm  id},\quad t(-4)=6\zeta\tilde{\zeta} \prod_{l=1}^N\rho_1
(-\theta_l)\,\times{\rm  id}, \no \\
&&t_2(0)=\frac{7}{2}
(1+c_1c_2-4\zeta^2)(1+\tilde{c}_1\tilde{c}_2-4\tilde{\zeta}^2)\prod_{l=1}^N\rho_{
\bar v}
(-\theta_l)\,\times{\rm   id},\no \\
&&t_2(-4)=\frac{7}{2}
(1+c_1c_2-4\zeta^2)(1+\tilde{c}_1\tilde{c}_2-4\tilde{\zeta}^2)\prod_{l=1}^N\rho_{
\bar v}
(-\theta_l)\,\times{\rm   id}, \no \\
 && t_2(-\frac{1}{2})= \frac{28}{3}\zeta\tilde{\zeta}\, t(-1),\quad t_2(-\frac{7}{2})= \frac{28}{3}\zeta\tilde{\zeta}
  \,t(-3),\no \\
&&t_3(0)=
2^7\cdot 7  \zeta\tilde{\zeta}(1+c_1c_2-4\zeta^2)(1+\tilde{c}_1\tilde{c}_2-4\tilde{\zeta}^2)
\prod_{l=1}^N\rho_{
\tilde v}
(-\theta_l)\times{\rm   id},\no \\
&&t_3(-4)=2^7\cdot 7 \zeta\tilde{\zeta}(1+c_1c_2-4\zeta^2)(1+\tilde{c}_1\tilde{c}_2-4\tilde{\zeta}^2)
\prod_{l=1}^N\rho_{
\tilde v}
(-\theta_l)\times{\rm   id}, \no \\
 && t_3(-1)= \frac{16\zeta\tilde{\zeta}}{\prod_{l=1}^N(1-\theta_l)(1+\theta_l)}t_2(-\frac{3}{2}),\quad
t_3(-3)=\frac{16\zeta\tilde{\zeta}}{\prod_{l=1}^N(1-\theta_l)(1+\theta_l)}\,t_2(-\frac{5}{2}),\no \\
 && t_3(-\frac{1}{2})= -\frac{28(1+c_1c_2-4\zeta^2)(1+\tilde{c}_1\tilde{c}_2-4\tilde{\zeta}^2)}
 {\prod_{l=1}^N(\frac{3}{2}-\theta_l)(\frac{3}{2}+\theta_l)}t(-\frac{3}{2}),\no\\
 && t_3(-\frac{7}{2})= -\frac{28(1+c_1c_2-4\zeta^2)(1+\tilde{c}_1\tilde{c}_2-4\tilde{\zeta}^2)}
 {\prod_{l=1}^N(\frac{3}{2}-\theta_l)(\frac{3}{2}+\theta_l)}t(-\frac{5}{2}).\label{t3-4}
 \eea
In the derivation, we have used the relations
\bea &&
tr \{K^{+}(0)\}=6\tilde{\zeta},\quad K^{-}(0)=\zeta\times {\rm id }, \quad tr\{K^{
-}(-4)\}=6\zeta,\quad K^{+}(-4)=\tilde{\zeta}\times {\rm id},\no\\[4pt]
&&tr \{K^{+}_{\bar 1} (0)\}=7(1+\tilde{c}_1\tilde{c}_2-4\tilde{\zeta}^2),\quad K^{-}_{\bar 1}(0)=\frac{1}{2}(1+c_1c_2-4\zeta^2)\times {\rm id },\no\\[4pt]
&&tr\{K^{-}_{\bar 1}(-4)\}=7(1+c_1c_2-4\zeta^2),\quad K^{+}_{\bar 1}(-4)=\frac{1}{2}(1+\tilde{c}_1\tilde{c}_2-4\tilde{\zeta}^2)\times {\rm id },\no\\[4pt]
&&tr \{K^{+}_{\tilde 1}(0)\}=2^4\cdot 7\tilde{\zeta}(1+\tilde{c}_1\tilde{c}_2-4\tilde{\zeta}^2),\quad K^{-}_{\tilde 1}(0)=8\zeta(1+c_1c_2-4\zeta^2)\times {\rm id },\no\\[4pt]
&&tr\{K^{-}_{\tilde 1}(-4)\}=2^4\cdot 7\zeta(1+c_1c_2-4\zeta^2),\quad K^{+}_{\tilde 1}(-4)=8\tilde{\zeta}(1+\tilde{c}_1\tilde{c}_2-4\tilde{\zeta}^2)\times {\rm id },\no\\[4pt]
&&tr_1\{R_{12}(-1)K^{+}_1(0)R_{21}(-7)\}=-2^4\cdot 3\cdot7\tilde{\zeta}\times {\rm id },\no\\[4pt]
&&tr_2\{R_{21}(-7)K^{-}_2(-4)R_{12}(-1)\}^{t_1t_2}=-2^4\cdot 3\cdot7 \zeta\times {\rm id },\no\\[4pt]
&&tr_1\{R_{12}(-1)R_{13}(-2)K^{+}_1(0)R_{31}(-6)R_{21}(-7)\}=
2^6\cdot 3^3\cdot 7\tilde{\zeta}\times {\rm id },\no\\[4pt]
 &&tr_3\{R_{31}(-6)R_{32}(-7)K^{-}_3(-4)R_{23}(-1)
 R_{13}(-2)\}^{t_1t_2t_3}=
2^6\cdot 3^3\cdot 7{\zeta}\times {\rm id },\no\\[4pt]
 && tr_{12}\{R_{23}(-6)R_{13}(-7)K^{ +}_2(-\frac{1}{2})R_{12}(-8)K^{+}_1(\frac{1}{2})P_{321} R_{32}(-1)R_{31}(-2)R_{21}(0)\}\no\\[4pt]
&&\quad\quad\quad\quad=2^8\cdot 3^4 \cdot7^2(1+\tilde{c}_1\tilde{c}_2-4\tilde{\zeta}^2)\times {\rm id },\no\\[4pt]
 && tr_{23}\{R_{21}(-6)R_{31}(-7)K^{ -}_2(-\frac{7}{2})R_{32}(-8)K^{
-}_1(-\frac{9}{2})P_{123} R_{12}(-1)R_{13}(-2)R_{23}(0)\}^{t_1t_2t_3}\no\\[4pt]
&&\quad\quad\quad\quad=2^8 \cdot3^4 \cdot7^2 (1+c_1c_2-4\zeta^2)\times {\rm id },\no\\[4pt]
&&K^{-}(-\frac{1}{2})K^{-}(\frac{1}{2})=\frac{1}{4}(1+c_1c_2-4\zeta^2)\times {\rm id
 },\;
K^{+}(-\frac{7}{2})K^{+}(-\frac{9}{2})=\frac{1}{4}(1+\tilde{c}_1\tilde{c}_2-4\tilde{\zeta}^2)\times
{\rm id }.\no \eea

From the construction of transfer matrices, we know that $t(u)$, $t_2(u)$ and $t_3(u)$ are the
operator polynomials of $u$ with degrees $4N+2$, $4N+4$ and $2N+6$, respectively.
Thus we need $10N+15$ independent conditions to determine their eigenvalues.

\subsection{Functional relations}

We have proved that the transfer matrices $t(u)$, $t_2(u)$ and $t_3(u)$
have common eigenstates. Acting the transfer matrices on the common eigenstates, we obtain the corresponding eigenvalues.
Denote the eigenvalues of
$t(u)$, $t_2(u)$ and $t_3(u)$ as $\Lambda(u)$,
$\Lambda_2(u)$ and $\Lambda_3(u)$, respectively.
Acting the operators (\ref{futpp-7}) on the common eigenstate, we obtain
that these eigenvalues satisfy following closed functional relations
\bea && \Lambda(\pm\theta_j)\Lambda(\pm\theta_j-4)={\frac{1}{2^4}}
\frac{(\pm\theta_j-\frac{3}{{2}})
(\pm\theta_j+\frac{3}{{2}})(\pm\theta_j-4)(\pm\theta_j+4)}
{(\pm\theta_j-\frac{1}{{2}})(\pm\theta_j+\frac{1}{{2}})
(\pm\theta_j-2)(\pm\theta_j+2)}\no\\[4pt]
&&\hspace{20mm}\times H_1(\pm\theta_j)H_2(\pm\theta_j)\varrho(\pm\theta_j)\varrho(\mp\theta_j), \no  \\[4pt]
&& \Lambda(\pm\theta_j)\Lambda(\pm\theta_j-1)= \frac{1}{ 2^{2}}
\frac{
(\pm\theta_j-1)(\pm\theta_j+\frac{3}{{2}})^2(\pm\theta_j+4)}
{(\pm\theta_j-\frac{1}{{2}})(\pm\theta_j+\frac{7}{{2}})
(\pm\theta_j+1)(\pm\theta_j+2)}\nonumber\\[4pt]
&&\hspace{20mm}\times \varrho(\pm\theta_j)\Lambda_2(\pm\theta_j-\frac{1}{{2}}), \no \\[4pt]
&& \Lambda(\pm\theta_j)\Lambda_2
(\pm\theta_j-\frac{3}{{2}})=\frac{1}{2^4}
\frac{(\pm\theta_j-\frac{3}{2})(\pm\theta_j+\frac{3}{{2}})
(\pm\theta_j+1)(\pm\theta_j+4)}{(\pm\theta_j-\frac{1}{{2}})(\pm\theta_j+\frac{1}{{2}})
(\pm\theta_j+2)(\pm\theta_j+3)}\nonumber\\[4pt]
&&\hspace{20mm}\times
\varrho(\pm\theta_j)\prod_{i=1}^N(\pm\theta_j-\theta_i+1)(\pm\theta_j+\theta_i+1)
\Lambda_3(\pm\theta_j-1), \no \\[4pt]
&&
\Lambda(\pm\theta_j)\Lambda_2(\pm\theta_j-\frac{7}{2})=\frac{1}{
2^{2}}
\frac{(\pm\theta_j-1)(\pm\theta_j-\frac{7}{{2}})(\pm\theta_j+\frac{3}{{2}})
(\pm\theta_j+4)}{(\pm\theta_j-\frac{1}{{2}})(\pm\theta_j-\frac{3}{{2}})(\pm\theta_j+1)
(\pm\theta_j+2)}\nonumber\\[4pt]
&&\hspace{20mm}\times
H_1(\pm\theta_j) H_2(\pm\theta_j)\varrho(\pm\theta_j)\Lambda(\pm\theta_j-3),\no \\[4pt]
&& \Lambda(\pm\theta_j)\Lambda_3(\pm\theta_j-3)=  \frac{
(\pm\theta_j-3)(\pm\theta_j+4)}{(\pm\theta_j-1)(\pm\theta_j+2) }
\prod_{i=1}^N(\pm\theta_j-\theta_i+4)(\pm\theta_j+\theta_i+4)\no\\
&&\hspace{20mm}\times H_1(\pm\theta_j)H_2(\pm\theta_j)\Lambda_2(\pm\theta_j-\frac{5}{2}), \quad j=1, \cdots, N.
\label{futpp-72sdf} \eea
The asymptotic behaviors (\ref{as-tv2}) imply
\bea && \Lambda(u)|_{u\rightarrow \pm\infty}=
-3(2+c_1\tilde{c}_2+c_2\tilde{c}_1)
 u^{4N+2}+\cdots, \no \\[4pt]
&& \Lambda_2(u)|_{u\rightarrow \pm\infty}=
2^{2}[3(2+c_1\tilde{c}_2+c_2\tilde{c}_1)^2+2(1+{c}_1{c}_2)
(1+\tilde{c}_1\tilde{c}_2)] u^{4N+4}+\cdots,\no \\[4pt]
&&\Lambda_3(u)|_{u\rightarrow \pm\infty}=-
2^{6}(2+c_1\tilde{c}_2+c_2\tilde{c}_1)[(2+c_1\tilde{c}_2+c_2\tilde{c}_1)^2+3(1+{c}_1{c}_2)(1+\tilde{c}_1\tilde{c}_2)]
u^{2N+6}\no\\&&\hspace{25mm}+\cdots. \label{as-tv21} \eea Besides,
from Eq.(\ref{t3-4}), we also have \bea
&&\Lambda(0)=6\zeta\tilde{\zeta} \prod_{l=1}^N\rho_1
(-\theta_l),\quad \Lambda(-4)=6\zeta\tilde{\zeta}
\prod_{l=1}^N\rho_1
(-\theta_l), \no \\
&&\Lambda_2(0)=\frac{7}{2}
(1+c_1c_2-4\zeta^2)(1+\tilde{c}_1\tilde{c}_2-4\tilde{\zeta}^2)\prod_{l=1}^N\rho_{
\bar v}
(-\theta_l),\no \\
&&\Lambda_2(-4)=\frac{7}{2}
(1+c_1c_2-4\zeta^2)(1+\tilde{c}_1\tilde{c}_2-4\tilde{\zeta}^2)\prod_{l=1}^N\rho_{
\bar v}
(-\theta_l), \no \\
 && \Lambda_2(-\frac{1}{2})= \frac{28}{3}\zeta\tilde{\zeta}\,\Lambda(-1),\quad \Lambda_2(-\frac{7}{2})= \frac{28}{3}\zeta\tilde{\zeta}
  \,\Lambda(-3),\no \\[4pt]
&&\Lambda_3(0)=
2^7\cdot 7 \zeta\tilde{\zeta}(1+c_1c_2-4\zeta^2)(1+\tilde{c}_1\tilde{c}_2-4\tilde{\zeta}^2)
\prod_{l=1}^N\rho_{ \tilde v}
(-\theta_l),\no \\
&&\Lambda_3(-4)=2^7\cdot 7 \zeta\tilde{\zeta}(1+c_1c_2-4\zeta^2)(1+\tilde{c}_1\tilde{c}_2-4\tilde{\zeta}^2)
\prod_{l=1}^N\rho_{ \tilde v}
(-\theta_l), \no \\
 && \Lambda_3(-1)= \frac{16\zeta\tilde{\zeta}}{\prod_{l=1}^N(1-\theta_l)(1+\theta_l)}
 \Lambda_2(-\frac{3}{2}),\;
\Lambda_3(-3)=\frac{16\zeta\tilde{\zeta}}{\prod_{l=1}^N(1-\theta_l)(1+\theta_l)}\,
\Lambda_2(-\frac{5}{2}),\no \\[6pt]
 && \Lambda_3(-\frac{1}{2})= -\frac{28(1+c_1c_2-4\zeta^2)(1+\tilde{c}_1\tilde{c}_2-
 4\tilde{\zeta}^2)}
 {\prod_{l=1}^N(\frac{3}{2}-\theta_l)(\frac{3}{2}+\theta_l)}\Lambda(-\frac{3}{2}),\no\\[6pt]
 && \Lambda_3(-\frac{7}{2})= -\frac{28(1+c_1c_2-4\zeta^2)(1+\tilde{c}_1
 \tilde{c}_2-4\tilde{\zeta}^2)}
 {\prod_{l=1}^N(\frac{3}{2}-\theta_l)(\frac{3}{2}+\theta_l)}\Lambda(-\frac{5}{2}).\label{t3-42sd}
 \eea
From above $10N$ functional relations (\ref{futpp-72sdf}),
3 asymptotic behaviors (\ref{as-tv21}) and 12 constraints (\ref{t3-42sd}),
we can completely determine the
eigenvalues $\Lambda(u)$, $\Lambda_2(u)$ and $\Lambda_3(u)$.

\subsection{Inhomogeneous $T-Q$ relations}

For the simplicity, we define some functions
\bea &&Z_1(u)=\frac{1}{ 2^{2}} \frac{
(u+\frac{3}{{2}})(u+4)}{(u+\frac{1}{{2}})(u+2) }A(u)\frac{Q^{(1)}(u-1)}{Q^{(1)}(u)} H_1(u),\no\\[4pt]
&&Z_2(u)=\frac{1}{ 2^{2}} \frac{
u(u+\frac{3}{{2}})(u+4)}{(u+\frac{1}{{2}})(u+1)(u+2)
}B(u)\frac{Q^{(1)}(u+1)Q^{(2)}(u-1)}{Q^{(1)}(u)Q^{(2)}(u)}H_2(u+1),\no\\[4pt]
&&Z_3(u)=\frac{1}{ 2^{2}}\frac{ u(u+4)}{(u+1)(u+2) }B(u)\frac{Q^{(2)}(u+1)Q^{(3)}(u-\frac{3}{2})}{Q^{(2)}(u)Q^{(3)}(u+\frac{1}{2})}H_1(u+1),\no\\[4pt]
&&Z_4(u)=\frac{1}{ 2^{2}}\frac{ u(u+4)}{(u+2)(u+3) }B(u)\frac{Q^{(2)}(u+1)Q^{(3)}(u+\frac{5}{2})}{Q^{(2)}(u+2)Q^{(3)}(u+\frac{1}{2})}H_2(u+3),\no\\[4pt]
&&Z_5(u)=\frac{1}{ 2^{2}}\frac{
u(u+\frac{5}{{2}})(u+4)}{(u+2)(u+3)(u+\frac{7}{{2}})
}B(u)\frac{Q^{(1)}(u+2)Q^{(2)}(u+3)}{Q^{(1)}(u+3)Q^{(2)}(u+2)}H_1(u+3),\no\\[4pt]
&&Z_6(u)=\frac{1}{ 2^{2}} \frac{
u(u+\frac{5}{{2}})}{(u+2)(u+\frac{7}{2}) }V(u)\frac{Q^{(1)}(u+4)}{Q^{(1)}(u+3)}H_2(u+4),\no\\[4pt]
&&f_1(u)=\frac{1}{ 2^{2}} \frac{u
(u+\frac{3}{{2}})(u+4)}{u+2 }B(u)G(u+1)\frac{Q^{(2)}(u-1)}{Q^{(1)}(u)}\,x,\no\\[4pt]
&&f_2(u)=\frac{1}{ 2^{2}}  u(u+4)B(u)\frac{Q^{(2)}(u+1)}{Q^{(3)}(u+\frac{1}{{2}})}\,x,\no\\[4pt]
&&f_3(u)=\frac{1}{ 2^{2}} \frac{ u(u+\frac{5}{{2}})(u+4)}{u+2
}B(u)G(u+3)\frac{Q^{(2)}(u+3)}{Q^{(1)}(u+3)}\, x, \label{function}
\eea where
$x=8\sqrt{(1+c_1c_2)(1+\tilde{c}_1\tilde{c}_2)}-4(2+c_1\tilde{c}_2+c_2\tilde{c}_1)$,
\bea
&& A(u)=\prod_{j=1}^N a(u-\theta_j)a(u+\theta_j), \quad B(u)=\prod_{j=1}^N b(u-\theta_j)b(u+\theta_j), \nonumber \\
&& V(u)=\prod_{j=1}^N e(u-\theta_j)e(u+\theta_j), \quad G(u)=\prod_{j=1}^N (u-\theta_j)(u+\theta_j),\no \\
&& Q^{(m)}(u)=\prod_{k=1}^{L_m}(u-\lambda_k^{(m)}+\frac{m}{2})(u+\lambda_k^{(m)}+\frac{m}{2}),
\quad m=1,2,3,
\eea
and the numbers of Bethe roots satisfy the constraints $L_1=L_2+N$ and $L_3=L_2$. By using these functions, we
construct the eigenvalues of transfer matrices as
\bea &&\Lambda(u)=\sum_{i=1}^6 \tilde{Z}_i(u),\no  \\[4pt]
&&\Lambda_2(u) =2^{-2}[(u-\frac12)(u+2)^2(u+\frac{9}{2})\varrho(u+\frac{1}{2})]^{-1}
\tilde{\rho}_{v}(2u) \no \\[4pt]
&&\qquad \times \left\{\sum^6_{i<j}
\tilde{Z}_i(u+\frac12)\tilde{Z}_j(u-\frac12)-\tilde{Z}_3(u+\frac12)\tilde{Z}_4(u-\frac12)\right.\no\\
&&\qquad \qquad \left.-f_1(u+\frac12)\tilde{Z}_2(u-\frac12)
-\tilde{Z}_5(u+\frac12)f_3(u-\frac12)\right\},\no  \\[4pt]
&& \Lambda_3(u)
=2^{-6}[(u+\frac{5}{{2}})^2(u+\frac{3}{{2}})^2(u-\frac{1}{{2}})(u+\frac{9}{{2}})
u(u-1)(u+2)^2(u+4)(u+5)\nonumber\\[4pt]
&&\qquad \times \varrho(u+1)\varrho(u)\prod_{i=1}^N
(u+\theta_i+2)(u-\theta_i+2)]^{-1}\tilde{\rho}_{ v}
(2u+1)\tilde{\rho}_{  v} (2u)\tilde{\rho}_{  v} (2u-1)
 \no\\[4pt]
&&\qquad
\times \left\{\sum^6_{i<j<k}\tilde{Z}_i(u+1)\tilde{Z}_j(u)\tilde{Z}_k(u-1)
-\sum^6_{k=5}\tilde{Z}_3(u+1)\tilde{Z}_4(u)\tilde{Z}_k(u-1) \right.\no \\
&&\qquad
\left.-\sum^2_{i=1}\tilde{Z}_i(u+1)\tilde{Z}_3(u)\tilde{Z}_4(u-1)
-\sum^4_{j=3}\tilde{Z}_2(u+1)\tilde{Z}_j(u)\tilde{Z}_5(u-1) \right.\no \\
&&\qquad
\left.
-\sum^6_{j=3}f_1(u+1)\tilde{Z}_2(u)\tilde{Z}_j(u-1)-\sum^4_{i=1}\tilde{Z}_i(u+1)\tilde{Z}_5(u)f_3(u-1)\right\},\label{eop-2321} \eea
where \bea&&\tilde{Z}_1(u)={Z}_1(u)+f_1(u),\quad
\tilde{Z}_2(u)={Z}_2(u),\no\\
&&\tilde{Z}_3(u)={Z}_3(u)+f_2(u),\quad
\tilde{Z}_4(u)={Z}_4(u),\no\\
&&\tilde{Z}_6(u)={Z}_6(u)+f_3(u),\quad \tilde{Z}_5(u)={Z}_5(u).\eea
All the eigenvalues are the polynomials, thus the residues of right
hand sides of Eq.(\ref{eop-2321}) should be zero, which gives the Bethe ansatz equations \bea &&\frac{1}
{\lambda_k^{(1)}(\lambda_k^{(1)}-\frac{1}{2})}\frac{h_1(\lambda_k^{(1)}-\frac{1}{2})
\tilde{h}_1(\lambda_k^{(1)}-\frac{1}{2})}{\prod_{j=1}^N
(\lambda_k^{(1)}-\theta_j-\frac{1}{2})(\lambda_k^{(1)}+\theta_j-\frac{1}{2})}
\frac{Q^{(1)}(\lambda_k^{(1)}-\frac{3}{2})}{Q^{(2)}(\lambda_k^{(1)}-\frac{3}{2})}
\no\\ && \qquad\qquad +\frac{1}
{\lambda_k^{(1)}(\lambda_k^{(1)}+\frac{1}{2})}\frac{h_2(\lambda_k^{(1)}+\frac{1}{2})
\tilde{h}_2(\lambda_k^{(1)}+\frac{1}{2})}{\prod_{j=1}^N
(\lambda_k^{(1)}-\theta_j+\frac{1}{2})(\lambda_k^{(1)}+\theta_j+\frac{1}{2})}
\frac{Q^{(1)}(\lambda_k^{(1)}+\frac{1}{2})}{Q^{(2)}(\lambda_k^{(1)}-\frac{1}{2})}=-x,
\no\\[6pt] && \qquad\qquad k=1, 2, \cdots, L_1, \no \\[6pt]
&&\frac{Q^{(1)}(\lambda_l^{(2)})Q^{(2)}(\lambda_l^{(2)}-2)Q^{(3)}(\lambda_l^{(2)}-\frac{1}{2})}
{Q^{(1)}(\lambda_l^{(2)}-1)Q^{(2)}(\lambda_l^{(2)})Q^{(3)}(\lambda_l^{(2)}-\frac{5}{2})}
\frac{h_2(\lambda_l^{(2)})\tilde{h}_2(\lambda_l^{(2)})}
{h_1(\lambda_l^{(2)})\tilde{h}_1(\lambda_l^{(2)})}=-\frac{\lambda_l^{(2)}-\frac{1}{2}}
{\lambda_l^{(2)}+\frac{1}{2}},\no\\[6pt]
&&\qquad\qquad l=1, 2,\cdots, L_2, \no \\[6pt]
 &&  \frac{h_1(\lambda_m^{(3)}-1)
\tilde{h}_1(\lambda_m^{(3)}-1)}{\lambda_m^{(3)}(\lambda_m^{(3)}-1)}
\frac{Q^{(3)}(\lambda_m^{(3)}-\frac{7}{2})}{Q^{(2)}(\lambda_m^{(3)}-2)}\no\\
&& \qquad\qquad+ \frac{h_2(\lambda_m^{(3)}+1)
\tilde{h}_2(\lambda_m^{(3)}+1)}{\lambda_m^{(3)}(\lambda_m^{(3)}+1)}
\frac{Q^{(3)}(\lambda_m^{(3)}+\frac{1}{2})}{Q^{(2)}(\lambda_m^{(3)})}=-x,\quad
 m=1, 2, \cdots, L_3. \label{opba-3} \eea
We note that from the regularity analysis of any $\Lambda(u)$,
$\Lambda_2(u)$ or $\Lambda_3(u)$, one can obtain the complete set
of Bethe ansatz equations. The Bethe ansatz equations obtained
from $\Lambda(u)$ are the same as those obtained from
$\Lambda_2(u)$ and $\Lambda_3(u)$. Meanwhile, the function
$Q^{(m)}(u)$ has two zero points, namely, $\lambda_k^{(m)}-\frac
m2$ and $-\lambda_k^{(m)}-\frac m2$. These two zero points should
give the same Bethe ansatz equations.

We have checked that the eigenvalues $\Lambda(u)$, $\Lambda_2(u)$ and $\Lambda_3(u)$ given by (\ref{eop-2321})
satisfy the closed fusion relations (\ref{futpp-72sdf}), asymptotic behaviors (\ref{as-tv21}) and constraints (\ref{t3-42sd}).
Therefore, we conclude that the eigenvalues constructed by the inhomogeneous $T-Q$ relations are indeed
the eigenvalues of transfer matrices, provided that the Bethe roots
satisfy Bethe ansatz equations (\ref{opba-3}). The eigenvalue
of Hamiltonian (\ref{hh}) can be expressed in terms of the Bethe
roots as
\begin{eqnarray}
E= \left.\frac{\partial \ln \Lambda(u)}{\partial
u}\right|_{u=0,\{\theta_j\}=0}.
\end{eqnarray}

If $c_1=c_2=\tilde{c}_1=\tilde{c}_2=0$, the boundary reflection matrices degenerate into the diagonal ones and our results cover
that obtained by the algebraic Bethe ansatz \cite{c2aba}.

\section{$C_{n}$ model}
\setcounter{equation}{0}

In this section, we generalize the above results to the $C_n$ model.
The $C_n$ model with periodic boundary condition has been studied
in reference \cite{Bn} by the algebraic Bethe ansatz method. Thus we focus on the open boundary
conditions. The $R$-matrix of the $C_{n}$ model is a
$(2n)^2\times (2n)^2$ one with the elements
\begin{eqnarray}
\bar R(u)^{ij}_{kl} = u(u+n+1)\delta_{ik}\delta_{jl}+
(u+n+1)\delta_{il}\delta_{jk}-u\xi_i\xi_k
\delta_{j\bar{i}}\delta_{k\bar{l}},
 \label{rmn}
\end{eqnarray}
where $i, j, k, l=1, \cdots, 2n$, $i+\bar{i}=2n+1$, $\xi_i=1$ if $i\in[1,n]$ and $\xi_i=-1$ if
$i\in[n+1,2n]$. The off-diagonal boundary reflection matrix $\bar K_0^{-}(u)$ is chosen as \bea \bar K_0^{-}(u)=\bar \zeta+\bar M_0 u, \quad
\bar M_0=\left(\begin{array}{cc}-1& \bar c_{1}\\[6pt]
    \bar c_{2}&1
 \end{array}\right)\otimes I,\label{K-matrix-VV-n}\eea
where $\bar \zeta$, $\bar c_1$ and $\bar c_2$ are the free boundary parameters and $I$ is a $n\times n$ unitary matrix. The dual reflection matrix
$\bar K_0^{+}(u)$ is determined by the mapping
\begin{equation}
\bar K_0^{ +}(u)=\bar K_0^{ -}(-u-n-1)|_{\bar \zeta, \bar c_i\rightarrow
    \tilde{\bar \zeta},\tilde{\bar c}_i }, \label{ksk2sd}
\end{equation}
where $\tilde{\bar \zeta}$, $\tilde{\bar c}_1$ and $\tilde{\bar c}_2$ are the
boundary parameters.

From $R$-matrix (\ref{rmn}) and reflection matrices (\ref{K-matrix-VV-n})-(\ref{ksk2sd}), the transfer matrix of $C_n$ model is constructed as
\begin{equation}
\bar t(u)= tr_0 \{\bar K_0^{+}(u)\bar T_0 (u) \bar K^{-}_0(u)\hat{\bar T}_0 (u)\}, \label{trweweucn}
\end{equation}
where
\begin{eqnarray}
&&\bar T_0(u)=\bar R_{01}(u-\theta_1)\bar R_{02}(u-\theta_2)\cdots \bar R_{0N}(u-\theta_N), \no \\
&&\hat{\bar T}_0 (u)=\bar R_{N0}(u+\theta_N)\cdots \bar R_{20}(u+\theta_{2}) \bar R_{10}(u+\theta_1).
\end{eqnarray}
The transfer matrix (\ref{trweweucn}) is the generating function of all the conserved quantities including the model Hamiltonian.

Now, we seek the eigenvalues of the transfer matrix (\ref{trweweucn}).
The main idea is similar as what we have done for the case of $n=3$.
For the generic $n$, we need $2n-1$ closed operator product identities. Besides the
\bea
\bar t_1(\theta_j)\,\bar t_1(\theta_j-n-1)&\sim& {\rm
id}, \label{fusion-1-Cn}
\eea  and $n-1$ relations (similar as those of the $A_n$ case \cite{Cao14})
\bea
\bar t_1(\theta_j)\,\bar t_m(\theta_j-\frac{m+1}{2})&\sim& \bar t_{m+1}(\theta_j-\frac m2), \quad m=1, 2, \cdots, n-1,\label{fusion-2-Cn}
 \eea
the other $n-1$ necessary relations are
 \bea
\bar t_1(\theta_j)\,\bar t_k(\theta_j-\frac{2n-k+3}{2})&\sim& \bar t_{k-1}(\theta_j-\frac{2n-k+2}{2}), \quad k=2, 3, \cdots, n.\label{fusion-3-Cn}
 \eea
Here, for simplicity we have ignored the coefficients in the above fusion relations.
The relations (\ref{fusion-1-Cn})-(\ref{fusion-3-Cn}) can also be demonstrated by the diagram
\bea
\xymatrix{
&\bar t_1\ar[r]^{\bar t_{1}}  &\bar t_2\ar[r]^{\bar t_{1}}    &\cdots\ar[r]^{\bar t_1}&\bar  t_{n-1}\ar[r]^{\bar t_1}&\bar t_{n}\ar[dl]^{\bar t_1}\\
 {\rm id}& \bar  t_1\ar[l]_{\bar t_{1}} &  \bar  t_2\ar[l]_{\bar t_1}& \cdots\ar[l]_{\bar t_1}
   &\bar t_{n-1}\ar[l]_{\bar t_{1}}&&}.
\label{fusion-4-Cn}
\eea
We note that in the diagram (\ref{fusion-4-Cn}), the values of spectral parameter in the fused transfer matrices are different.
The above $2n-1$ fusion relations (\ref{fusion-1-Cn})-(\ref{fusion-3-Cn}), together with the associated
asymptotic behaviors (similar as those (\ref{as-tv2}) for the $n=3$ case) and the special values of transfer matrices at certain points (similar as those (\ref{t3-4}) for the $n=3$ case),
allow us to construct the inhomogeneous $T-Q$ relations of all the fused transfer matrices. Here we  present the finial results.

The eigenvalues of the transfer matrix (\ref{trweweucn}) can be given by
\bea
&&\bar \Lambda(u)=\sum_{l=1}^{2n}\bar Z_l(u)+\sum_{j=1}^{n} \bar
f_j(u).\label{Eigen-open-Lambda-n}\eea The functions $\bar
Z_l(u)$ in Eq.(\ref{Eigen-open-Lambda-n}) are defined as \bea &&\bar Z_1(u)=\frac{1}{ 2^{2}} \frac{
(u+\frac{n}{{2}})(u+n+1)}{(u+\frac{1}{{2}})(u+\frac{n+1}{{2}})
}\bar A(u)\frac{\bar Q^{(1)}(u-1)}{\bar Q^{(1)}(u)}\bar H_1(u),\no\\[4pt]
&&\bar Z_{2n}(u)=\frac{1}{ 2^{2}} \frac{
u(u+\frac{n+2}{{2}})}{(u+\frac{n+1}{{2}})(u+n+\frac{1}{2})
}\bar V(u)
\frac{\bar Q^{(1)}(u+n+1)}{\bar Q^{(1)}(u+n)}\bar H_{2n}(u+n+1), \no \\[4pt]
&&\bar Z_l(u)=\frac{1}{ 2^{2}} \frac{
u(u+\frac{n}{{2}})(u+n+1)}{(u+\frac{l-1}{{2}})(u+\frac{l}{{2}})(u+\frac{n+1}{{2}})
}\bar B(u)\frac{\bar Q^{(l-1)}(u+1)\bar Q^{(l)}(u-1)}{\bar Q^{(l-1)}(u)\bar Q^{(l)}(u)} \bar H_l(u),\no\\[4pt]
&&\bar Z_{2n-l+1}(u)=\frac{1}{ 2^{2}} \frac{
u(u+\frac{n}{{2}}+1)(u+n+1)}{(u+n-\frac{l-2}{{2}})(u+n-\frac{l-3}{{2}})(u+\frac{n+1}{{2}})
}\bar B(u)\no\\[4pt]
&&\hspace{10mm}\times \frac{\bar Q^{(l-1)}(u+n-l+1)\bar Q^{(l)}(u+n-l+2)}{\bar Q^{(l-1)}
(u+n-l+2)\bar Q^{(l)}(u+n-l+1)} \bar H_{2n-l+1}(u),\quad l=2,3,\cdots,n-1, \no\\[4pt]
&&\bar Z_n(u)=\frac{1}{ 2^{2}}\frac{
u(u+n+1)}{(u+\frac{n+1}{2})(u+\frac{n-1}{2}) }\bar B(u)\frac{\bar Q^{(n-1)}(u+1)\bar Q^{(n)}(u-\frac{3}{2})}{\bar Q^{(n-1)}(u)\bar Q^{(n)}(u+\frac{1}{2})}\bar H_n(u),\no\\[4pt]
&&\bar Z_{n+1}(u)=\frac{1}{ 2^{2}}\frac{
u(u+n+1)}{(u+\frac{n+1}{2})(u+\frac{n+3}{2}) }\bar B(u)\frac{\bar Q^{(n-1)}(u+1)\bar Q^{(n)}(u+\frac{5}{2})}
{\bar Q^{(n-1)}(u+2)\bar Q^{(n)}(u+\frac{1}{2})}\bar H_{n+1}(u), \eea
where
\bea
&& \bar A(u)=\prod_{j=1}^N (u-\theta_j+1)(u-\theta_j+n+1)(u+\theta_j+1)(u+\theta_j+n+1), \no\\
&& \bar B(u)=\prod_{j=1}^N (u-\theta_j)(u-\theta_j+n+1)(u+\theta_j)(u+\theta_j+n+1), \nonumber \\
&& \bar V(u)=\prod_{j=1}^N (u-\theta_j)(u-\theta_j+n)(u+\theta_j)(u+\theta_j+n), \no \\
&& \bar Q^{(m)}(u)=\prod_{k=1}^{\bar L_m}(u-\lambda_k^{(m)}+\frac{m}{2})(u+\lambda_k^{(m)}+\frac{m}{2}),
\quad m=1,2,\cdots,n, \no\\[4pt]
&&\bar H_l(u)=\left\{\begin{array}{cc}
\bar h_1(u+\frac {l-1}{2}),&l \in odd \; \; in\; \; [1, n], \\[6pt]
\bar h_2(u+\frac l2),& l \in even \;\; in \; \; [1, n],
 \end{array}\right.\no\\[6pt]
&&\bar H_{2n-l+1}(u)=\left\{\begin{array}{cc}
\bar h_2(u+n+1-\frac{l-1}{2}),&l \in odd \;\; in \;\; [1, n],\\[6pt]
\bar h_1(u+n+1-\frac l2),& l \in even \; \;in \;\; [1, n],
 \end{array}\right.\no\\[6pt]
&& \bar h_1(u)=-4(\sqrt{(1+\bar c_1\bar c_2)}u+\bar \zeta)(\sqrt{1+\tilde{\bar c}_1\tilde{\bar c}_2}u-\tilde{\bar \zeta}),\no \\[4pt]
&& \bar h_2(u)=-4(\sqrt{(1+\bar c_1\bar c_2)}u-\bar
\zeta)(\sqrt{1+\tilde{\bar c}_1\tilde{\bar c}_2}u+\tilde{\bar
\zeta}), \label{no} \eea and the numbers of Bethe roots satisfy
the constraints \bea \bar L_1=\bar L_2+N,\quad \bar L_{2l-1}=\bar
L_{2l-2}+\bar L_{2l},\quad \bar L_n=\bar L_{n-1},\quad
l=2,3,\cdots,\frac{n-1}{2},\eea if $n$ is odd and the constraints
\bea \bar L_1=\bar L_2+N,\quad \bar L_{2l-1}=\bar L_{2l-2}+\bar
L_{2l},\quad \bar L_{n-1}=\bar L_{n-2}+2\bar L_n,\quad
l=2,3,\cdots,\frac{n-2}{2},\eea if $n$ is even. The inhomogeneous
terms $\bar f_i(u)$ with odd $n$ are also different from that with
even $n$. If $n$ is odd, we have \bea &&\bar f_1(u)=\frac{1}{
2^{2}} \frac{u
(u+\frac{n}{{2}})(u+n+1)}{u+\frac{n+1}{{2}} }\bar B(u) \bar G(u+1)\frac{\bar Q^{(2)}(u-1)}{\bar Q^{(1)}(u)}\bar x,\no\\[4pt]
&&\bar f_n(u)=\frac{1}{ 2^{2}} \frac{
u(u+\frac{n}{{2}}+1)(u+n+1)}{u+\frac{n+1}{{2}} }\bar B(u)\bar G(u+n)
\frac{\bar Q^{(2)}(u+n)}{\bar Q^{(1)}(u+n)}\bar  x,\no \\
&&\bar f_l(u)=\frac{1}{ 2^{2}} \frac{u
(u+\frac{n}{{2}})(u+n+1)}{u+\frac{n+1}{{2}} }\bar B(u)
\frac{\bar Q^{(2l-2)}(u+1)\bar Q^{(2l)}(u-1)}{\bar Q^{(2l-1)}(u)}\bar x, \no\\[4pt]
&&\bar f_{n-l+1}(u)=\frac{1}{ 2^{2}} \frac{
u(u+\frac{n}{{2}}+1)(u+n+1)}{u+\frac{n+1}{{2}} }\bar B(u)
\no\\[4pt]
&&\hspace{10mm}\times
\frac{\bar Q^{(2l)}(u+n+2-2l)\bar Q^{(2l-2)}(u+n+2-2l)}{\bar Q^{(2l-1)}(u+n+2-2l)}\bar x,\quad
l=2,\cdots,[\frac{n}{{2}}]-1, \label{no1} \eea
and
\bea&&\bar f_{\frac{n+1}{{2}}}(u)=\frac{1}{ 2^{2}}
u(u+n+1)\bar B(u)
\frac{\bar Q^{(n-1)}(u+1)}{\bar Q^{(n)}(u+\frac{1}{{2}})}\bar x,\label{no2} \eea
where $\bar x=8\sqrt{(1+\bar c_1\bar c_2)(1+\tilde{\bar c}_1\tilde{\bar c}_2)}-4(2+\bar c_1\tilde{\bar c}_2+\bar c_2\tilde{\bar c}_1)$ and
\bea \bar  G(u)=\prod_{j=1}^N (u-\theta_j)(u+\theta_j).
\eea
If $n$ is even, besides the $n-2$ inhomogeneous terms
$\bar f_i(u)$ given by (\ref{no1}), the rest two read \bea&&\bar f_{\frac{n}{{2}}}(u)=\frac{u
(u+\frac{n}{{2}})(u+n+1)}{u+\frac{n+1}{{2}} }\bar B(u)
\frac{\bar Q^{(n-2)}(u+1)\bar Q^{(n)}(u-\frac{1}{{2}})\bar Q^{(n)}(u-\frac{3}{{2}})}{\bar Q^{(n-1)}(u)}\bar x,\no\\[4pt]
&&\bar f_{\frac{n}{{2}}+1}(u)=\frac{u
(u+\frac{n}{{2}}+1)(u+n+1)}{u+\frac{n+1}{{2}} }\bar B(u)
\no\\[4pt]
&&\hspace{20mm}\times
\frac{\bar Q^{(n-2)}(u+2)\bar Q^{(n)}(u+\frac{3}{{2}})\bar Q^{(n)}(u+\frac{5}{{2}})}{\bar Q^{(n-1)}(u+2)}\bar x.\label{no3}
\eea
We note that if $n=2$, $\bar Q^{(0)}(u)=\bar G(u)$, $\bar L_0=N$, the functions $\bar f_1(u)$ and $\bar f_2(u)$ are defined by
Eq.(\ref{no3}) instead of (\ref{no1}) because of the present parametrization.

From the singularities analysis of inhomogeneous $T-Q$ relations (\ref{Eigen-open-Lambda-n}), we obtain the Bethe ansatz equations, which also
depend on the parity of $n$. If $n$ is odd, the Bethe ansatz equations are
\bea &&\frac{1}
{\lambda_k^{(1)}(\lambda_k^{(1)}-\frac{1}{2})}\frac{\bar h_1(\lambda_k^{(1)}-\frac{1}{2})}{\prod_{j=1}^N
(\lambda_k^{(1)}-\theta_j-\frac{1}{2})(\lambda_k^{(1)}+\theta_j-\frac{1}{2})}
\frac{\bar Q^{(1)}(\lambda_k^{(1)}-\frac{3}{2})}{\bar Q^{(2)}(\lambda_k^{(1)}-\frac{3}{2})}
\no\\ && \qquad\quad +\frac{1}
{\lambda_k^{(1)}(\lambda_k^{(1)}+\frac{1}{2})}\frac{\bar h_2(\lambda_k^{(1)}+\frac{1}{2})
}{\prod_{j=1}^N
(\lambda_k^{(1)}-\theta_j+\frac{1}{2})(\lambda_k^{(1)}+\theta_j+\frac{1}{2})}
\frac{\bar Q^{(1)}(\lambda_k^{(1)}+\frac{1}{2})}{\bar Q^{(2)}(\lambda_k^{(1)}-\frac{1}{2})}=-\bar x,
\no\\[8pt] && \qquad\quad k=1,2,\cdots, \bar L_1, \no \\[8pt]
&&\frac{\bar h_1(\lambda_k^{(l)}-\frac{1}{2})
}{\lambda_k^{(l)}(\lambda_k^{(l)}-\frac{1}{2})}
\frac{\bar Q^{(l)}(\lambda_k^{(l)}-\frac{l}{2}-1)}{\bar Q^{(l-1)}(\lambda_k^{(l)}-\frac{l}{2})
\bar Q^{(l+1)}(\lambda_k^{(l)}-\frac{l}{2}-1)} \no\\ &&
\qquad\quad
+\frac{\bar h_2(\lambda_k^{(l)}+\frac{1}{2})
}{\lambda_k^{(l)}(\lambda_k^{(l)}+\frac{1}{2})}
\frac{\bar Q^{(l)}(\lambda_k^{(l)}-\frac{l}{2}+1)}{\bar Q^{(l-1)}(\lambda_k^{(l)}-\frac{l}{2}+1)
\bar Q^{(l+1)}(\lambda_k^{(l)}-\frac{l}{2})}=-\bar x,
\no\\[8pt] && \qquad\quad k=1,2,\cdots, \bar  L_l, \qquad\qquad  l \in odd\; \; in \;\; [2, n-2], \no \\[8pt]
&&\frac{\bar Q^{(l-1)}(\lambda_k^{(l)}-\frac{l}{2}+1)\bar Q^{(l)}(\lambda_k^{(l)}-\frac{l}{2}-1)
\bar Q^{(l+1)}(\lambda_k^{(l)}-\frac{l}{2})}
{\bar Q^{(l-1)}(\lambda_k^{(l)}-\frac{l}{2})\bar Q^{(l)}(\lambda_k^{(l)}-\frac{l}{2}+1)
\bar Q^{(l+1)}(\lambda_k^{(l)}-\frac{l}{2}-1)}
\frac{\bar h_2(\lambda_k^{(l)})}
{\bar h_1(\lambda_k^{(l)})}=-\frac{\lambda_k^{(l)}-\frac{1}{2}}
{\lambda_k^{(l)}+\frac{1}{2}},\no\\[8pt]
&&\qquad\quad k=1,2,\cdots, \bar L_l,  \qquad \qquad l \in even\; \; in \;\; [2, n-2], \label{opba-2-1}
\eea
and the rest two read
\bea
&&\frac{\bar Q^{(n-2)}(\lambda_k^{(n-1)}-\frac{n-1}{2}+1)\bar Q^{(n-1)}(\lambda_k^{(n-1)}-\frac{n-1}{2}-1)
\bar Q^{(n)}(\lambda_k^{(n-1)}-\frac{n-1}{2}+\frac 12)}
{\bar Q^{(n-2)}(\lambda_k^{(n-1)}-\frac{n-1}{2})\bar Q^{(n-1)}(\lambda_k^{(n-1)}-\frac{n-1}{2}+1)
\bar Q^{(n)}(\lambda_k^{(n-1)}-\frac{n-1}{2}-\frac 32)}\no\\
&&\qquad \qquad \times
\frac{\bar h_2(\lambda_k^{(n-1)})}
{\bar h_1(\lambda_k^{(n-1)})}=-\frac{\lambda_k^{(n-1)}-\frac{1}{2}}
{\lambda_k^{(n-1)}+\frac{1}{2}}, \qquad k=1,2,\cdots, \bar L_{n-1},\no \\
&&\frac{\bar h_1(\lambda_k^{(n)}-1)
}{\lambda_k^{(n)}(\lambda_k^{(n)}-1)}
\frac{\bar Q^{(n)}(\lambda_k^{(n)}-\frac{n+1}{2}-\frac
32)}{\bar Q^{(n-1)}(\lambda_k^{(n)}-\frac{n+1}{2}) }
+\frac{\bar h_2(\lambda_k^{(n)}+1)
}{\lambda_k^{(n)}(\lambda_k^{(n)}+1)}
\frac{\bar Q^{(n)}(\lambda_k^{(n)}-\frac{n+1}{2}+\frac
52)}{\bar Q^{(n-1)}(\lambda_k^{(n)}-\frac{n+1}{2}+2) }=-\bar x,
\no\\[4pt]
&&\qquad \qquad
k=1,2,\cdots,\bar L_n. \label{opba-1-n-2} \eea
If $n$ is even, besides (\ref{opba-2-1}), the rest two are
\bea&&\frac{\bar h_1(\lambda_k^{(n-1)}-\frac{1}{2})
}{\lambda_k^{(n-1)}(\lambda_k^{(n-1)}-\frac{1}{2})}
\frac{\bar Q^{(n-1)}(\lambda_k^{(n-1)}-\frac{n-1}{2}-1)}{\bar Q^{(n-2)}(\lambda_k^{(n-1)}-\frac{n-1}{2})
\bar Q^{(n)}(\lambda_k^{(n-1)}-\frac{n-1}{2}-\frac 32)} \no\\
&& \qquad \qquad
+\frac{\bar h_2(\lambda_k^{(n-1)}+\frac{1}{2})
}{\lambda_k^{(n-1)}(\lambda_k^{(n-1)}+\frac{1}{2})}
\frac{\bar Q^{(n-1)}(\lambda_k^{(n-1)}-\frac{n-1}{2}+1)}{\bar Q^{(n-2)}(\lambda_k^{(n-1)}-\frac{n-1}{2}+1)
\bar Q^{(n)}(\lambda_k^{(n-1)}-\frac{n-1}{2}+\frac
12)}\no\\
&& \qquad \qquad=-\bar x\bar Q^{(n)}(\lambda_k^{(n-1)}-\frac{n-1}{2}-\frac 12), \quad k=1,2,\cdots, \bar L_{n-1}, \label{BAEs} \\[8pt]
&&\frac{\bar h_2(\lambda_k^{(n)}-\frac 12)
}{\lambda_k^{(n)}(\lambda_k^{(n)}-1)}
\frac{\bar Q^{(n)}(\lambda_k^{(n)}-\frac{n+1}{2}-\frac
32)}{\bar Q^{(n-1)}(\lambda_k^{(n)}-\frac{n+1}{2}) }
+\frac{\bar h_1(\lambda_k^{(n)}+\frac 12)
}{\lambda_k^{(n)}(\lambda_k^{(n)}+1)}
\frac{\bar Q^{(n)}(\lambda_k^{(n)}-\frac{n+1}{2}+\frac
52)}{\bar Q^{(n-1)}(\lambda_k^{(n)}-\frac{n+1}{2}+2) }=0,
\no\\[4pt]
&& \qquad \qquad k=1,2,\cdots, \bar L_n. \label{opba-1-n-2-1} \eea
We note that if $n=2$, the Bethe roots are determined by
Eqs.(\ref{BAEs}) and (\ref{opba-1-n-2-1}) due to the
parametrization we used\footnote{We note that the results (\ref{Eigen-open-Lambda-n})-(\ref{opba-1-n-2-1}) coincide with
those of \cite{Li_219} for the case of $n=2$ because all the eigenvalues and Bethe ansatz equations are derived from the same fusion relations which completely determine the eigenvalue functions.
The present forms (\ref{Eigen-open-Lambda-n})-(\ref{opba-1-n-2-1}) and those in \cite{Li_219} are actually equivalent but with different parametrizations.
If $n=3$, Eqs.(\ref{Eigen-open-Lambda-n})-(\ref{opba-1-n-2-1}) are reduced to (\ref{eop-2321})-(\ref{opba-3}).}.
\begin{table}[!htbp]
\caption{Numerical solutions for the case of $n=2$ and $N=2$. Here $u=0.3$, the boundary parameters are chosen as
$c_{1}=c_{2}=0$, $\tilde{c}_1= 0.3$, $\tilde{c}_2=0.1$, $\zeta= 0.2 $, $\tilde{\zeta}=0.4$, and ``${\rm deg}$" means the degeneracy.
We see that the BAEs (\ref{BAEs}) and (\ref{opba-1-n-2-1}) can give all the 16 levels of eigenvalues of the transfer matrix (\ref{trweweucn}). }
\begin{center}
{\footnotesize
\begin{tabular}{c c c c c c c c}
\hline \hline n &deg & $\bar \Lambda_{n}$ & $\lambda^{(1)}_{1}$ & $\lambda^{(1)}_{2}$ & $\lambda^{(1)}_{3}$ & $\lambda^{(1)}_{4}$ & $\lambda^{(2)}_{1}$ \\
  \hline1 &1 & -24.8089 & 0.9876 & 2.118 & 20.86 & 0.2920 & 1.087 \\
  \hline2 &1 & -12.1340 & 3.635+11.41i & 0.2948 & 3.635-11.41i & 20.76 & 8.063i \\
  \hline3 &1 &  14.3524 & 20.18 & 11.63 & 7.060i  & 0.5041i & 0.8461 \\
  \hline4 &1& 198.9426 & 3.649-11.36i & 0.9210 & 20.77 & 3.649+11.36i & 7.983i \\
  \hline5 &1 & -73.2660 & 18.84-3.661i & 18.84+3.661i & 0.6156 & 0.3902 & 16.10 \\
  \hline6 &1 & -11.0547 & 1.164 & 17.50+2.590i & 0.3003 & 17.50-2.590i & 0.4694 \\
  \hline7 &1 &  45.6395 & 11.60i & 13.04+9.083i & -13.04+9.083i & 22.02 & 12.40 \\
  \hline8 &1 & 289.3355 & 0.9004 & 17.53+2.601i & 17.53-2.601i & 1.602 & 1.198 \\
   \hline9  &2& -73.1354 & 0.6154 & 0.3903 & -& - &-  \\
   \hline10 &2 & -12.1318 & 0.2948 & 12.76 & -& - &-\\
   \hline11 &2 &  45.4078 & 7.357i & 10.35& -& - &-\\
   \hline12 &2& 200.8723 & 0.9210 & 12.78& -& - &-\\
   \hline\hline
\end{tabular}
 }
\end{center}
\end{table}

\begin{table}[!htbp]
%\begin{sidewaystable}[!htbp]
\caption{Numerical solutions for the case of $n=2$ and $N=3$. Here $u=0.3$, the boundary parameters are chosen as
$c_{1}=c_{2}=0$, $\tilde{c}_1= 0.3$, $\tilde{c}_2=0.1$, $\zeta= 0.2 $, $\tilde{\zeta}=0.4$, and ``${\rm deg}$" means the degeneracy.
We see that the BAEs (\ref{BAEs}) and (\ref{opba-1-n-2-1}) can give all the 64 levels of eigenvalues of the transfer matrix (\ref{trweweucn}). }
\begin{center}
{%\footnotesize
%\scriptsize
\tiny
\begin{tabular}{cccc cccc c}
\hline\hline n&deg &$\bar \Lambda_{n}$  & $\lambda^{(1)}_{1}$ & $\lambda^{(1)}_{2}$ & $\lambda^{(1)}_{3}$ & $\lambda^{(1)}_{4}$ & $\lambda^{(1)}_{5}$& $\lambda^{(2)}_{1}$  \\
   \hline1&2&-1309.0722086& 0.8137 - 0.1513i   &0.2952   &20.1674 - 2.7688i   & 0.8137 + 0.1513i  &-20.1674 - 2.7688i  &17.7865\\
  \hline2 &2&-770.9540442&21.7348 - 6.1393i  &21.7348 + 6.1393i   &0.2995 &-0.9636 + 0.0593i   &0.9636 + 0.0593i  &  - 0.9590i\\
  \hline3 &2&-746.6466105&-1.3561 &-0.8674  & 2.3847  &23.8959   &0.3006   &1.3824\\
  \hline4&2& -522.489193& 0.3012 & 2.6417 +13.7871i   &2.6417 -13.7871i  &23.7861 &-0.8432&11.4780i\\
  \hline5&2 &-452.8625437&21.7534 - 6.1363i   &1.9901  &21.7534 + 6.1363i   &0.8965 &0.2997  &0.9326 \\
  \hline6&2&-224.3281084& 0.9096  &23.9099 & 0.7478i   &1.9749   &0.2994   &0.9687 \\
  \hline7 &2&-134.104649&21.8273 - 6.5476i  &-1.2742  &21.8273 + 6.5476i    &0.3000  &-6.4817 &-0.5763\\
  \hline8 &2&-126.476708& 3.2002i  &21.7231 - 6.0035i  &21.7231 + 6.0035i   &1.2649  &0.3000 & 0.5274 \\
  \hline9&2 &-87.9711782&-3.7168 +16.8398i  &22.8394   &0.2997   &-12.9128  &3.7168 +16.8398i  & -13.0794i\\
  \hline10&2&-80.1865787& 1.8385i   &2.5937 +13.6641i   &2.5937 -13.6641i  &23.8028  &0.2996   & +11.4162i\\
  \hline11&2&-34.4681321& 0.5568i   &5.8157  &21.8022 - 6.4601i   &0.2997   &21.8022 + 6.4601i  &- 0.6602i \\
  \hline12&2& -33.1089124& 21.7353 - 6.0972i  &-0.2995 &21.7353 + 6.0972i   &0.3780 - 0.7062i  &-0.3780 - 0.7062i &  - 0.3852i\\
  \hline13&2&150.1067751& 21.1838 &5.1648 +10.1978i  & 5.1648 -10.1978i   & 0.2891i  &18.9727 & 0.7784\\
  \hline14&2&264.019449& 21.8083 - 6.4404i  &-0.6965  &21.8083 + 6.4404i  &-5.7246   &0.3032 &0.7003 \\
  \hline15&2 &277.9657081&-12.8373 & 0.4911i  & 3.7066 +16.8207i &22.8532  &3.7066 -16.8207i   &13.0760i \\
  \hline16&2&481.7478967&-5.6216&-0.5378 + 0.8564i  &21.7931 - 6.4396i &21.7931 + 6.4396i   &0.5378 + 0.8564i  &- 0.9564i \\
  \hline17&2&503.6398593&-5.1888 +10.1494i  &19.0899&0.8699i  &21.0860  &5.1888 +10.1494i  &-1.0254  \\
  \hline18&2&883.857325&-17.8344 -11.2725i  & 6.0052 +15.9459i &-17.8344 +11.2725i &26.3666  &-6.0052 +15.9459i &-14.0555 \\
  \hline19&2 &1427.2505815&21.7495 + 6.1237i  &-0.8968  & 0.4253i   &21.7495 - 6.1237i   &1.6685  &0.8486  \\
  \hline20&2&1697.8214136&21.8081 + 6.4493i  &21.8081 - 6.4493i  &-5.7824 &- 0.5860i   &0.8925 &-0.5563\\
  \hline21&2&3895.5472569& 23.8246& 2.5238 +13.4919i & 3.1256i  &-0.8967  &-2.5238 +13.4919i   &11.3331i\\
  \hline22&2&4179.3844963& 22.8106 &-3.7379 +16.8785i & -3.7379 -16.8785i  &0.8960  &13.0687   &13.0852i\\
  \hline23&2&5021.6412418&  0.8944 &-21.7288 - 5.9582i &-21.7288 + 5.9582i  &-1.7786 & 3.8063i   &1.2319  \\
  \hline24&2&5253.7513642&  1.8108  &21.8522 + 6.6002i  &21.8522 - 6.6002i   &6.8947 &0.8943 &-1.3211i   \\
      \hline25&2&-1307.7494557 &0.8134 + 0.1513i  &-0.8134 + 0.1513i   &0.2951 + 0.0000i&-&-&-\\
  \hline26&2& -524.3995776 &-0.3012 + 0.0000i  &-0.8436 - 0.0000i  &17.2846 - 0.0000i&-&-&-\\
  \hline27&2&-89.055869  &5.7396 + 0.0000i  &17.1999 + 0.0000i   &0.2997 + 0.0000i&-&-&-\\
  \hline28&2&-79.6620626  &0.2996 - 0.0000i  &-0.0000 - 1.7202i &-17.2942 - 0.0000i&-&-&-\\
  \hline29&2&279.844251 &-5.4653 + 0.0000i  &-0.0000 - 0.4845i  &17.2098 - 0.0000i&-&-&-\\
  \hline30&2&882.6244989 &-15.2683 + 0.0000i  &-5.4235 +10.2884i  &-5.4235 -10.2884i&-&-&-\\
  \hline31&2&3840.6635657 &17.3038 + 0.0000i   &0.8969 - 0.0000i   &0.0000 - 2.6931i&-&-&-\\
  \hline32&2&4254.1367807 &0.8958 - 0.0000i  &-17.1810 - 0.0000i   &6.2013 - 0.0000i&-&-&-\\
  \hline  \hline
\end{tabular}
 }
\end{center}
%\end{sidewaystable}
\end{table}

Now we check above analytic results by the numerical calculations.
For simplicity, we consider the case of $n=2$ with system-size $N=2,3$ and random choices of the boundary parameters.
We first solve the Bethe ansatz equations (BAEs) (\ref{BAEs}) and (\ref{opba-1-n-2-1}) and obtain
the values of Bethe roots. Substituting these Bethe roots into the inhomogeneous $T-Q$ relation (\ref{Eigen-open-Lambda-n}), we obtain
the corresponding eigenvalues of the transfer matrix (\ref{trweweucn}). The results are given in Table I ($N=2$) and Table II ($N=3$), where the boundary parameters are chosen as
$c_{1}=c_{2}=0$, $\tilde{c}_1= 0.3$, $\tilde{c}_2=0.1$, $\zeta= 0.2 $, $\tilde{\zeta}=0.4$ and the
spectral parameter $u$ is put as $u=0.3$.
We also diagonalize the transfer matrix (\ref{trweweucn}) by using the numerical exact diagonalization method.
We find that the analytical results and the numerical results are consistent with each other very well.
Thus the BAEs (\ref{BAEs}) and (\ref{opba-1-n-2-1}) can give the complete solutions of the corresponding model.

\section{Discussion}

In this paper, we study the exact solutions of the $C_n$ vertex model with either the periodic or the open boundary conditions corresponding to the
$K$-matrices (\ref{K-matrix-VV-n})-(\ref{ksk2sd})
by using fusion and the nested off-diagonal Bethe ansatz. Taking the $C_3$ model as an example, we obtain its fusion structures and provide a way to close the recursive operator product identities among the transfer matrices. Based on them and some necessary additional information such as the
asymptotic behaviors and the relations at some special points, we obtain the eigenvalues (\ref{eop-2321}) of the system and give the associated Bethe ansatz equations (\ref{opba-3}).
Moreover, we also generalize these results (\ref{Eigen-open-Lambda-n})-(\ref{opba-1-n-2-1}) to the $C_n$ model with off-diagonal boundary reflections (\ref{K-matrix-VV-n})-(\ref{ksk2sd}).
The method and results given in this paper can be generalized to other high rank quantum integrable systems.

\section*{Acknowledgments}

We would like to thank Prof. Y. Wang for his valuable discussions
and continuous encouragement.
The financial supports from National Program for Basic Research of
MOST (Grant Nos. 2016 YFA0300600 and 2016YFA0302104), National
Natural Science Foundation of China (Grant Nos. 11934015,
11975183, 12047502, 12075177, 12074410, 11947301, 11774397, 11775178 and 11775177),
Major Basic Research Program of Natural Science of Shaanxi
Province (Grant Nos. 2017KCT-12, 2017ZDJC-32), Australian Research
Council (Grant No. DP 190101529), Strategic Priority Research
Program of the Chinese Academy of Sciences (Grant No. XDB33000000), and Double First-Class
University Construction Project of Northwest University are
gratefully acknowledged.

%%%%%%%%%%%%%%%%%%%%%%%%%%%%%%%%%%%%%%%%%%%%%%%%%%%%%%%%%%%%%%%
%                                                             %
%  References                                                 %
%                                                             %
%%%%%%%%%%%%%%%%%%%%%%%%%%%%%%%%%%%%%%%%%%%%%%%%%%%%%%%%%%%%%%%

\end{document}